\documentclass[a4paper,11pt]{revtex4}

\usepackage[T1]{fontenc} 
\usepackage{amsmath}
\usepackage{graphicx}

\newcommand{\be}{\begin{eqnarray}}
\newcommand{\ee}{\end{eqnarray}}

 \newcommand{\gsim}{\mathrel{\hbox{\rlap{\lower.55ex \hbox {$\sim$}}
                   \kern-.3em \raise.4ex \hbox{$>$}}}}
\newcommand{\lsim}{\mathrel{\hbox{\rlap{\lower.55ex \hbox {$\sim$}}
                   \kern-.3em \raise.4ex \hbox{$<$}}}}

\newcommand{\ba}{\begin{eqnarray}}
\newcommand{\ea}{\end{eqnarray}}

\begin{document}
\title{Anomalous Chiral Transport in Heavy Ion Collisions from Anomalous-Viscous Fluid Dynamics} 
\author{Shuzhe Shi} \email{shishuz@indiana.edu}
\address{Physics Department and Center for Exploration of Energy and Matter,
Indiana University, 2401 N Milo B. Sampson Lane, Bloomington, IN 47408, USA.}

\author{Yin Jiang} 
\address{School of Physics and Nuclear Energy Engineering, Beihang University, Beijing 100191, China.}
\address{Institut f\"ur Theoretische Physik, Universit\"at Heidelberg, Philosophenweg 16, D-69120 Heidelberg, Germany.}

\author{Elias Lilleskov} 
\address{Department of Physics and Astronomy, Macalester College, 
1600 Grand Avenue, Saint Paul, MN 55105, USA.}
\address{Physics Department and Center for Exploration of Energy and Matter,
Indiana University, 2401 N Milo B. Sampson Lane, Bloomington, IN 47408, USA.}

\author{Jinfeng Liao} \email{liaoji@indiana.edu}
\address{Physics Department and Center for Exploration of Energy and Matter,
Indiana University, 2401 N Milo B. Sampson Lane, Bloomington, IN 47408, USA.}
\address{Institute of Particle Physics and Key Laboratory of Quark \& Lepton Physics (MOE), Central China Normal University, Wuhan, 430079, China.}

\date{\today}

\begin{abstract}
{  
Chiral anomaly is a fundamental aspect of quantum theories with chiral fermions. How such microscopic anomaly manifests itself in a macroscopic many-body system with chiral fermions, is a highly nontrivial question that has recently attracted significant interest. As it turns out, unusual transport currents can be induced by chiral anomaly under suitable conditions in such systems, with the notable example of the Chiral Magnetic Effect (CME) where a vector current (e.g. electric current) is generated along an external magnetic field. A lot of efforts have been made to search for  CME in heavy ion collisions, by measuring the charge separation effect induced by the CME transport. A crucial challenge in such effort, is the quantitative prediction for the CME signal. In this paper, we develop the Anomalous-Viscous Fluid Dynamics (AVFD) framework, which implements the anomalous fluid dynamics to describe the evolution of fermion currents in QGP, on top of the neutral bulk background described by the VISH2+1 hydrodynamic simulations for heavy ion collisions. With this new tool, we quantitatively and systematically investigate the dependence of the CME signal to a series of theoretical inputs and associated uncertainties. With realistic estimates of initial conditions and magnetic field lifetime, the predicted CME signal is quantitatively consistent with measured change separation data in 200GeV Au-Au collisions. Based on analysis of Au-Au collisions, we further make predictions for the CME observable to be measured in the planned isobaric (Ru-Ru v.s. Zr-Zr ) collision experiment, which could provide a most decisive test of the CME in heavy ion collisions.
}
\end{abstract}

\maketitle

\section{Introduction}

Symmetry principles play instrumental roles in the construction of our most basic physical theories. A special category of ``symmetry'' is the so-called anomaly, which is a well-defined classical symmetry of a theory but gets broken at quantum level. A most famous example of anomaly is the chiral anomaly  which is a very fundamental aspect of  quantum theories with spin-$\frac{1}{2}$ chiral fermion, from the Standard Model to supersymmetric field theories or even string theories. In such theories, the classical conservation law for the right-handed(RH) or left-handed(LH) chiral current $J^\mu_{R/L}$ gets broken at quantum level when coupled to (Abelian or non-Abelian) gauge fields.  The famous Adler-Bell-Jackiw anomaly for one species of  chiral fermion (with electric charge $Q_f$) in electromagnetic fields is given by~\cite{Adler:1969gk,Bell:1969ts}
\begin{eqnarray} \label{eq_CA}
\partial_\mu J^\mu_{R/L} = \pm \, C_A   \vec{\bf E} \cdot \vec{\bf B}
\end{eqnarray}
where $C_A=\frac{Q_f^2}{4\pi^2}$ is a universal coefficient. In the case of non-Abelian anomaly, one simply replaces the electromagnetic fields by appropriate non-Abelian gauge fields with slight modification of the constant $C_A$.  

Microscopic symmetry principles also manifest themselves nontrivially in macroscopic physics. For example, the fluid dynamics as a general long-time large-distance effective description of any macroscopic system, is a  direct manifestation of symmetries in the underlying microscopic dynamics. The fluid dynamical equations for energy-momentum tensor and for charged currents are the direct consequences of conservation of energy, momentum and charge which all originate from corresponding microscopic symmetries.

This naturally brings up a deep question: what are the implications of microscopic quantum anomaly (as a sort of ``half symmetry'') on the macroscopic properties of matter? Such question has triggered  significant interest and important progress recently.    As it turns out, unusual macroscopic transport currents can be induced by chiral anomaly   under suitable  conditions , with the notable example of the Chiral Magnetic Effect (CME) where a vector current (e.g. electric current $\vec{\bf J}$) is generated along an external magnetic field $\vec{\bf B}$~\cite{Vilenkin:1980fu,Kharzeev:2004ey,Kharzeev:2007tn,Kharzeev:2007jp,Fukushima:2008xe}:
\begin{eqnarray} \label{eq_CME}
\vec{\bf J}  =  C_A   \mu_5 \vec{\bf B} 
\end{eqnarray}
where the quantity $\mu_5$ is a chiral chemical potential that quantifies the macroscopic imbalance between RH and LH fermions in the system. Most remarkably, the conducting coefficient of this current $C_A$, being the  as the anomaly coefficient $C_A$ in Eq.(\ref{eq_CA}), is totally dictated by the microscopic anomaly relation.  The other highly nontrivial feature of the CME is that  the   anomalous transport process underlying this quantum current in the above equation is time-reversal even, i.e. non-dissipative~\cite{Kharzeev:2011ds}.

In the context of strong interaction physics as described by Quantum Chromodynamics (QCD), the chiral anomaly provides a unique access to the topological configurations such as instantons and sphelarons which are known to play crucial roles in many nonperturbative  phenomena~\cite{Schafer:1996wv,Diakonov:2009jq}. In particular, accompanying such topological configurations are the fluctuations of chirality imbalance (i.e. difference in the number of RH versus LH quarks) in the system, precisely due to the anomaly relation. That is, the macroscopic chirality fluctuations of fermions in the system, which would be experimentally measurable, reflect directly the gluonic topological fluctuations the information about which would be otherwise unaccessible. Furthermore, the CME (\ref{eq_CME}) provides a nontrivial way to manifest a nonzero macroscopic chirality (as quantified by $\mu_5$) in QCD matter: a nonzero $\mu_5$ with the presence of an external magnetic field $\vec{\bf B}$ would be measurable via the induced electric current $\vec{\bf J}$.

 The experimental realization of the CME has been enthusiastically pursued in two very different types of real world materials. The first is  the Dirac and Weyl semimetals in condensed matter physics~\cite{Nielsen:1983rb,Son:2012bg,Li:2014bha,Xiong:2015nna, 2015PhRvX...5c1023H,2016NatCo...711615A} where the CME has been successfully observed. The other is  the quark-gluon plasma (QGP) in heavy ion collision experiments at the Relativistic Heavy Ion Collider (RHIC) and the Large Hadron Collider (LHC)~\cite{STAR_LPV1,STAR_LPV2,STAR_LPV3,STAR_LPV4,STAR_LPV_BES,ALICE_LPV,Khachatryan:2016got}: see recent reviews in e.g. \cite{Kharzeev:2015znc,Liao:2014ava,Bzdak:2012ia,Huang:2015oca}. Encouraging evidence of CME-induced charge separation signals in those collisions have been reported, albeit with ambiguity due to background contamination. Crucial for addressing such issue, is the need of quantitative predictions for CME signals with sophisticated modelings. As a crucial step toward achieving this goal, we develop the Anomalous-Viscous Fluid Dynamics (AVFD) framework~\cite{Jiang:2016wve}, which implements the fluid dynamical evolution of chiral fermion currents (of the light flavor quarks) with anomalous transport in QGP on top of the expanding (neutral) bulk background described by the VISH2+1 hydrodynamic simulations. With this newly developed tool, we report our systematic and quantitative investigations on the CME signals in heavy ion collisions as well as the influence of various existing theoretical uncertainties.

The rest of the paper is organized as follows. 
In Sec.~\ref{sec.observable} a brief discussion on the relevant experimental observables, which are also the quantities to be computed, will be given. 
The detailed implementation of the  Anomalous-Viscous Fluid Dynamics (AVFD) framework will be described 
in Sec.~\ref{sec.framework}. Then the Sec.~\ref{sec.result} presents the quantitative results for the CME signals in  200GeV Au-Au collisions at RHIC, and the   Sec.~\ref{sec.isobar} focuses on the prediction for CME signals in the upcoming isobaric collision experiments. In addition, we will explore and quantify the influence of a few important theoretical uncertainties, including: the possible thermal relaxation effect on anomalous currents (in Sec.~\ref{sec.relaxative}); the finite mass and strangeness transport (in Sec.~\ref{sec.strangeness}); as well as possible out-of-equilibrium CME-induced charge separation from pre-hydro stage (in Sec.~\ref{sec.prehydro}). Finally the conclusion will be given in Sec.~\ref{sec.conclusion}.

\section{Charge Separation Measurements in Heavy Ion Collisions}\label{sec.observable}

In relativistic heavy-ion collisions the nuclei are accelerated to travel at nearly the speed of light. The strong Lorentz boost effect changes the nuclear geometry as observed in the laboratory frame, and more importantly, leads to a very strong magnetic field, which is Lorentz-transformed from the Coulomb-like electric field in the nucleus's co-moving frame. In a typical off-central  collisions, a magnetic field along the out-of-plane direction (conventionally defined as the $\hat{\boldsymbol{y}}$-direction) is  in the overlap region, where the hot quark-gluon plasma (QGP) also forms. For a given collision event with chirality imbalance (e.g. say, with more RH than LH light quarks corresponding to a positive chiral chemical potential $\mu_5>0$), the CME-induced electric current $\vec{\bf J}$ as in Eq.~(\ref{eq_CA}) is expected and it transports positive-charged particles toward the direction of the magnetic field while negative-charge particles toward the opposite direction. Similarly for events with more LH particles, i.e. a negative $\mu_5<0$, the CME current $\vec{\bf J}$ flips its direction, thus transporting positive- / negative-charged particles oppositely to the $\mu_5>0$ case.

Such CME-induced charge transport will lead to accumulation of excessive positive- / negative- charges above and below the reaction plane, i.e. a charge dipole moment along the out-of-plane (or equivalently the $\vec{\bf B}$ field) direction. This charge separation effect, when carried by the bulk collective flow, will lead to a specific pattern of   the azimuthal distribution of the finally observed charged hadrons particles in the momentum space, as follows:
\be
\frac{dN^\pm}{d\phi} \propto 1 + 2a_1^{\pm} \sin(\phi-\Psi_{RP}) + 2v_2 \cos(2\phi-2\Psi_{RP}) + ...
\label{eq_chargedistribution}
\ee
Here, $\Psi_{RP}$ represents the azimuthal angle of reaction plane, coefficient $v_2$ is the elliptic flow, while the CME charge separation is represented by the dipole term $a_1^{\pm} \sin(\phi-\Psi_{RP}) $ with $a_1^+ = - a_1^-$.

Note however  that the direction of the CME current flips with the chirality imbalance (arising from fluctuations), and there are equal probabilities for the event-wise chirality imbalance to be positive or negative. Therefore the event-averaged measurement of $a_1^\pm$ is expected to be vanishing, i.e. $\left< a_1^\pm \right> =0$. Clearly one can only measure the variance of such charged dipole. This can be done by measuring  the azimuthal correlations for same-charge and opposite-charge hadron pairs:
\be
\gamma_{\alpha\beta} \equiv \left< \cos(\phi_i+\phi_j-2\Psi_{RP}) \right>_{\alpha\beta},\qquad\qquad
\delta_{\alpha\beta} \equiv \left< \cos(\phi_i-\phi_j) \right>_{\alpha\beta},
\ee
where $\alpha,\beta=+$ or $-$ representing positive- / negative-charged particles. In the absence of {\sl ``true''} two-particle correlations, i.e. with the two-particle distribution of the form:
\be
f(\phi_i,\phi_j) \equiv f(\phi_i) f(\phi_j) + C(\phi_i,\phi_j) \to   f(\phi_i) f(\phi_j) ,
\ee
 one can derive that
\be
\gamma_{\alpha\beta}^{\mathrm{CME}} = - \left< a_{1,\alpha} a_{1,\beta}\right>,\qquad\qquad
\delta_{\alpha\beta}^{\mathrm{CME}} = \left< a_{1,\alpha} a_{1,\beta}\right>.
\ee 
It thus seems that by measuring $\gamma_{++,--}$ versus $\gamma_{+-}$, one could extract the CME-induced  signal. This is however naive, and turns out not working well due to the presence of substantial background correlations in the neglected $C(\phi_i,\phi_j)$ term in the above. Indeed a number of analyses clearly demonstrated that  the correlators $\gamma, \delta$ are strongly influenced by non-CME, flow-driven background contributions, such as the local charge conservation \cite{Schlichting:2010qia,Pratt:2010zn} and the transverse momentum conservation \cite{Bzdak:2009fc,Bzdak:2010fd,Liao:2010nv}, etc. 
If one adopts a two-component model analysis as developed in \cite{Bzdak:2012ia,Bloczynski:2013mca},  then the correlators could be decomposed into the background contribution $F$, and the ``pure'' CME signal $H\equiv\left< a_{1,\alpha} a_{1,\beta} \right>$. They contribute differently into the $\gamma$ and $\delta$ correlators, and in particular the flow-driven background would contribute to the correlators as $\delta^{\mathrm{bkg}} = F$ and $\gamma^{\mathrm{bkg}}=\kappa v_2 F$.  We therefore obtain the following decomposition relations: 
\be
\gamma=\kappa v_2 F - H, \qquad\qquad \delta = F + H.
\label{eq_cme_correlation}
\ee
Here the factor $\kappa$ quantifies the amount of $v_2$-driven background, which is expected to be in the range of $1\sim1.5$ and the AMPT simulation gives  the expectation that $\kappa\sim1.2$ (see e.g. \cite{Wen:2016zic}).

In the present paper we will focus on event-averaged smooth hydro simulations for quantifying the CME-induced signals $H_{SS} \equiv H_{++,--} = (a^{ch}_1)^2$ and $H_{OS} \equiv H_{+-} = -(a^{ch}_1)^2$, to be compared with the extracted signals by STAR Collaboration. A full investigation of the $\gamma$ and $\delta$ correlations would necessarily require event-by-event AVFD simulations that include fluctuations, hadronic cascades as well as various background contributions, which we leave as a future goal of study to be reported elsewhere.  

\section{The Anomalous-Viscous Fluid Dynamics (AVFD) Framework}\label{sec.framework}

To quantitatively study the  signal of CME, one needs to properly describe the transport of the light fermions (as RH and LH particles) in the hydrodynamic framework and to account for anomaly. Furthermore the  environment  created in a high-energy heavy ion collision is not a static medium but dominantly an almost-neutral hot fireball undergoing a strong collective expansion. Here we adopt a linearization approach to treat the RH and LH fermion currents as perturbations on top of the expanding bulk matter, with the following evolution equations:  
\begin{eqnarray}
\hat{D}_\mu J_{f,R}^\mu &=& + \frac{N_c Q_f^2}{4\pi^2} E_\mu B^\mu, \label{eq.Rcurrent}\\
\hat{D}_\mu J_{f,L}^\mu &=& - \frac{N_c Q_f^2}{4\pi^2} E_\mu B^\mu. \label{eq.Lcurrent}
\end{eqnarray}
These equations are solved as a linear perturbation on top of the neutral viscous fluid background, which is described by boost-invariant VISH2+1 hydrodynamic simulation~\cite{Shen:2014vra}. The VISH2+1 provides an excellent description of the bulk evolution and the computed collective flow observables agree well with a large body of available data.  
Adopted from VISH2+1, we employ the coordinates $(\tau,x,y,\eta)$, converting the Minkowski space-time coordinates $z-t$ into proper time $\tau\equiv\sqrt{t^2-z^2}$ and rapidity $\eta\equiv\frac{1}{2}\ln\frac{t+z}{t-z}$, with metric convention $g_{\mu\nu}=(1,-1,-1,-\tau^2)$.
We denote the projection operator $\Delta^{\mu \nu}=\left(g^{\mu\nu} - u^\mu u^\nu \right)$ where $u^\mu$ is the fluid velocity field, 
while  $\hat{d} =u^\mu \hat{D}_\mu$ with $\hat{D}_\mu$ as the covariant derivatives. 
In this coordinate system with non-zero affine connections 
$\Gamma^\rho_{\mu\nu}\equiv\frac{1}{2}g^{\rho\sigma}(\partial_{\nu}g_{\sigma\mu}+\partial_{\mu}g_{\sigma\nu}-\partial_{\sigma}g_{\mu\nu})$, the covariant derivative acting on a Lorentz scaler $S$, vector $V^\mu$ and tensor $W^{\mu\nu}$ can be expressed as
\begin{eqnarray}
\hat{D}_\mu S = \partial_\mu S, \quad
\hat{D}_\mu V^\nu = \partial_\mu V^\nu + \Gamma^{\nu}_{\lambda\mu}V^\lambda, \quad
\hat{D}_\mu W^{\nu\rho} = \partial_\mu W^{\nu\rho} + \Gamma^{\nu}_{\lambda\mu} W^{\lambda\rho} + \Gamma^{\rho}_{\lambda\mu} W^{\nu\lambda}.
\end{eqnarray}

It is worth mentioning that when including non-zero vector/axial charge $n = \lambda\,s$, one could expect that thermal quantities like $\epsilon$, $p$, $s$ and $T$ would be modified by $\sim\lambda^2$. 
The influence on hydro background from the back-reaction of nonzero net charge densities is rather small for 200GeV collisions, and thus the linearized approach here provides a very good approximate description. 
However, the influence of finite charge becomes crucial for collisions at low beam energy, where net vector/axial charges become substantial.
To study anomalous transport in low energy collisions, one should in principle solve the full dynamic equations coupling current transport (Eqs.\ref{eq.Rcurrent}\&\ref{eq.Lcurrent}) with evolution of the energy-momentum tensor.

\subsection{Viscous Fluid Dynamical Description of Heavy Ion Collisions}\label{subsec.vishnu}

VISH2+1 hydro package~\cite{Shen:2014vra} is an open source hydrodynamics code package developed by the Ohio State University group. 
It describes the evolution of the QGP, created by relativistic heavy-ion collisions, by assuming the system to be boost-invariant and charge neutral,  both of which are valid for ultra-relativistic heavy-ion collisions (e.g. for top energy collisions at RHIC as well as collisions at the LHC). 
It adopts the Isreal-Steward framework for the second-order viscous hydrodynamic equations, with the fluid energy momentum tensor given by: 
\be
T^{\mu\nu} = \varepsilon \,u^\mu u^\nu - (p+\Pi)\,\Delta^{\mu\nu} + \pi^{\mu\nu}.
\ee 
Implementing the lattice-based equation of state {s95p-v0-PCE} \cite{Huovinen:2009yb}, with the neutrality assumption   $n_f\equiv0$, 
VISH2+1 hydro solves the energy-momentum conservation equation
\be \label{eq_bulk}
\hat{D}_\nu T^{\mu\nu} = 0,\label{eq_IShydro_0}
\ee
with the relaxation equation of the bulk pressure $\Pi$ and shear stress tensor $\pi^{\mu\nu}$ toward the corresponding Navier-Stokes form,
\begin{eqnarray}
\Delta^{\mu\alpha}\Delta^{\nu\beta}\hat{d} \pi_{\alpha\beta} &=& 
-\frac{1}{\tau_\pi}( \pi^{\mu\nu} - 2\eta \sigma^{\mu\nu} )
- \frac{\pi^{\mu\nu}}{2} \frac{\eta T}{\tau_\pi} \hat{D}_\lambda\left(\frac{\tau_\pi}{\eta T} u^\lambda \right)
\label{eq_IShydro_1}\\
\hat{d} \Pi &=& -\frac{1}{\tau_\Pi}(\Pi+\zeta \theta) - \frac{\Pi}{2} \frac{\zeta T}{\tau_\Pi} \hat{D}_\lambda\left(\frac{\tau_\Pi}{\zeta T} u^\lambda \right) \,\, .
\label{eq_IShydro_2}
\end{eqnarray}
 After specifying suitable initial conditions,   one can then obtain the spatial distribution as well as time evolution of the temperature, energy density, pressure, as well as the fluid velocity of the QGP by solving the above differential equations. 
At the end of the fluid evolution,   the QGP hadronizes at a specific temperature, the freeze-out temperature $T_f$, 
and the final hadrons are then locally produced   in all fluid cells on the freeze-out hyper-surface with a  local thermal-equilibrium distribution including viscous corrections (see e.g. \cite{Shen:2014vra} for details), following the Cooper-Frye freeze-out formula 
\begin{eqnarray}\label{eq_cooperfrye}
E \frac{dN}{d^3p} (x^\mu, p^\mu) = \frac{g}{(2\pi)^3} \int_{\Sigma_{\rm fo}} p^\mu d^3\sigma_\mu f(x,p) \,\, .
\end{eqnarray} 
It is worth mentioning that during the hadronic stage, the hadron scatterings  and resonance decay processes would modify the finally observed hadronic spectra. This has been properly implemented in the VISH2+1 simulations and the results provide good agreement with measurements of  identified hadron observables.  
Finally, concerning the initial conditions of the bulk evolution, we use the averaged smooth initial conditions (for single-shot hydro evolution) from event-by-event Monte-Carlo Glauber model as implemented in VISH2+1 package. Again we emphasize that all these choices are following the standard VISH2+1 setup which has been successfully validated with various experimental data for soft bulk observables.

\subsection{Fermion Currents and Anomalous Chiral Transport}\label{subsec.avfd}

Let us focus on the collision with relatively high beam energy (such as the top energy collision at RHIC with $\sqrt{s_{NN}}=200\rm GeV$). The matter produced at such energy has rather small net conserved charges as compared with bulk energy or entropy and its evolution is usually well described by viscous hydrodynamics assuming neutrality for all fermion currents (as is the case for the VISH2+1). However in order to describe the charge transport in QGP, one needs to include the corresponding fluid dynamical evolution for the fermions (i.e. quarks and antiquarks which carry all the conserved charges). These currents though could be treated in a perturbative way, i.e. by evolving them on top of the neutral bulk fluid background (as specified by the space-time dependent temperature field $T(x^\mu)$ and fluid velocity field $u^\nu(x^\mu)$ from solving Eqs.(\ref{eq_bulk})) and ignoring their back reaction to the bulk evolution. Furthermore one would like to implement the anomalous transport effect in the fluid dynamics framework~\cite{Son:2009tf}.  The corresponding fluid dynamical equations for the  evolution of both RH and LH fermion currents for each light flavor of quarks, take the following form: 
\begin{eqnarray} \label{eq_avfd} 
\hat{D}_\mu J_{\chi, f}^\mu &=& \chi \frac{N_c Q_f^2}{4\pi^2} E_\mu B^\mu \\
J_{\chi, f}^\mu &=& n_{\chi, f}\, u^\mu + \nu_{\chi, f}^\mu + \chi \frac{N_c Q_f}{4\pi^2} \mu_{\chi, f} B^\mu 
\label{eq_avfd_J}   \\ 
\Delta^{\mu}_{\,\, \nu} \hat{d} \left(\nu_{\chi, f}^\nu \right) &=& - \frac{1}{\tau_{r}} \left[  \left( \nu_{\chi, f}^\mu \right) -  \left(\nu_{\chi, f}^\mu \right)_{NS} \right ] 
\label{eq_avfd_ns_1}\\
\left(\nu_{\chi, f}^\mu \right)_{NS} &=&  \frac{\sigma}{2} T \Delta^{\mu\nu}   \partial_\nu \left(\frac{\mu_{\chi, f}}{T}\right) +  \frac{\sigma}{2} Q_f E^\mu   \quad
\label{eq_avfd_ns_2}
\end{eqnarray} 
where $\chi=\pm1$ labels chirality for RH/LH currents and $f=u,d$ labels different light quark flavors with their respective electric charge $Q_f$ and with color factor $N_c=3$. The $E^\mu=F^{\mu\nu}u_\nu$ and $B^\mu=\frac{1}{2}\epsilon^{\mu\nu\alpha\beta}u_\nu F_{\alpha\beta}$ represent the external electromagnetic fields in fluid's local rest frame. 
Furthermore the (small) fermion densities $n_{\chi, f}$ and corresponding chemical potential $ \mu_{\chi, f}$ are related by lattice-computed quark number susceptibilities $c_2^f(T)$\cite{Borsanyi:2011sw}. It is worth emphasizing that the above framework treats the normal viscous currents $\nu^\mu_{\chi, f}$ at the second-order of gradient expansion by incorporating relaxation  toward Navier-Stokes form (which is the first-order gradient term), thus in consistency with the background bulk flow which is also described by the 2nd-order viscous hydrodynamics as shown in Eq. (\ref{eq_IShydro_0}-\ref{eq_IShydro_2}). Two key transport coefficients (characterizing normal viscous effects) are explicitly involved: the normal diffusion coefficient $\sigma$ and the relaxation time $\tau_r$. 

It should be particularly emphasized that the term $\chi \frac{N_c Q_f}{4\pi^2} \mu_{\chi, f} B^\mu$ in the Eq.(\ref{eq_avfd_J}) implements explicitly the CME current. Its sign changes with the chirality $\chi$, which reflects the feature of anomalous transport where the direction of the CME current is opposite for RH and LH particles.  
Note also that in the above framework, the CME current is treated as first-order gradient expansion term without any second-order thermal relaxation effect, owing to the argument that the CME current is of quantum nature. Nevertheless this is still an open question and a certain fraction of the CME current may still suffer from relaxation effect (see e.g. recent discussions in \cite{Kharzeev:2016sut,Huang:2017tsq}). In Sec.~\ref{sec.relaxative} we will introduce the second-order relaxation term for the CME current in a phenomenological way and investigate the potential influence on the CME signal  by such relaxation effect.

By solving  these equations with given initial conditions, one can obtain detailed information on the space-time evolution of the fermion currents  of any flavor or chirality  which accounts for both normal viscous charge transport and the anomalous transport. One can then determine the relevant chemical potential for each species of hadrons on the hydrodynamic freeze-out hyper-surface and include such nonzero chemical potential in  the Cooper-Frye formula~(\ref{eq_cooperfrye}). After including necessary hadron cascade processes, especially the contributions of resonance decay, one can obtain the momentum distribution of the hadrons which would be eventually measured by detectors.

\subsection{Comparison of Normal and Anomalous Transport}

\begin{figure*}[!hbt]
\begin{center}
\includegraphics[width=0.23\textwidth]{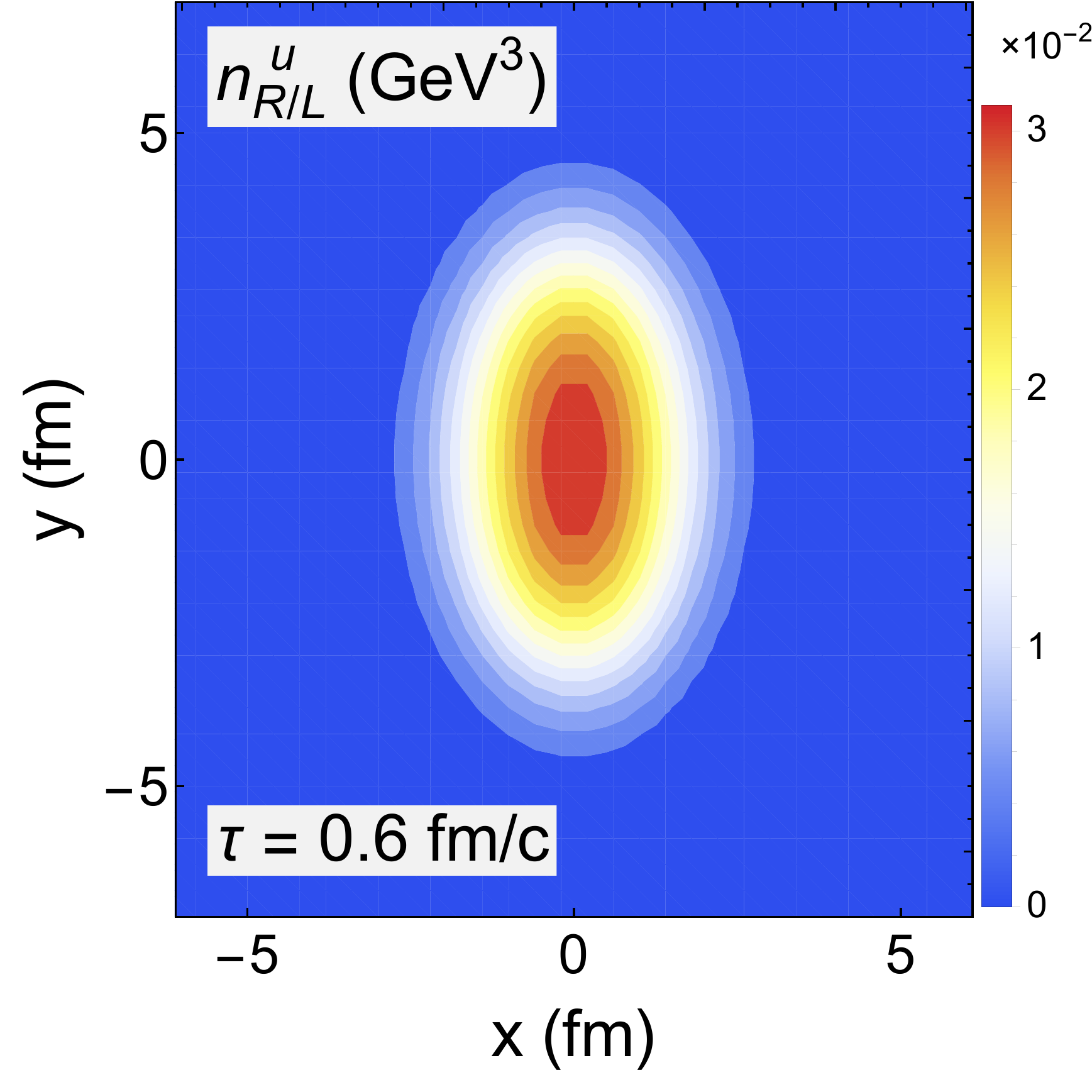} 
\includegraphics[width=0.23\textwidth]{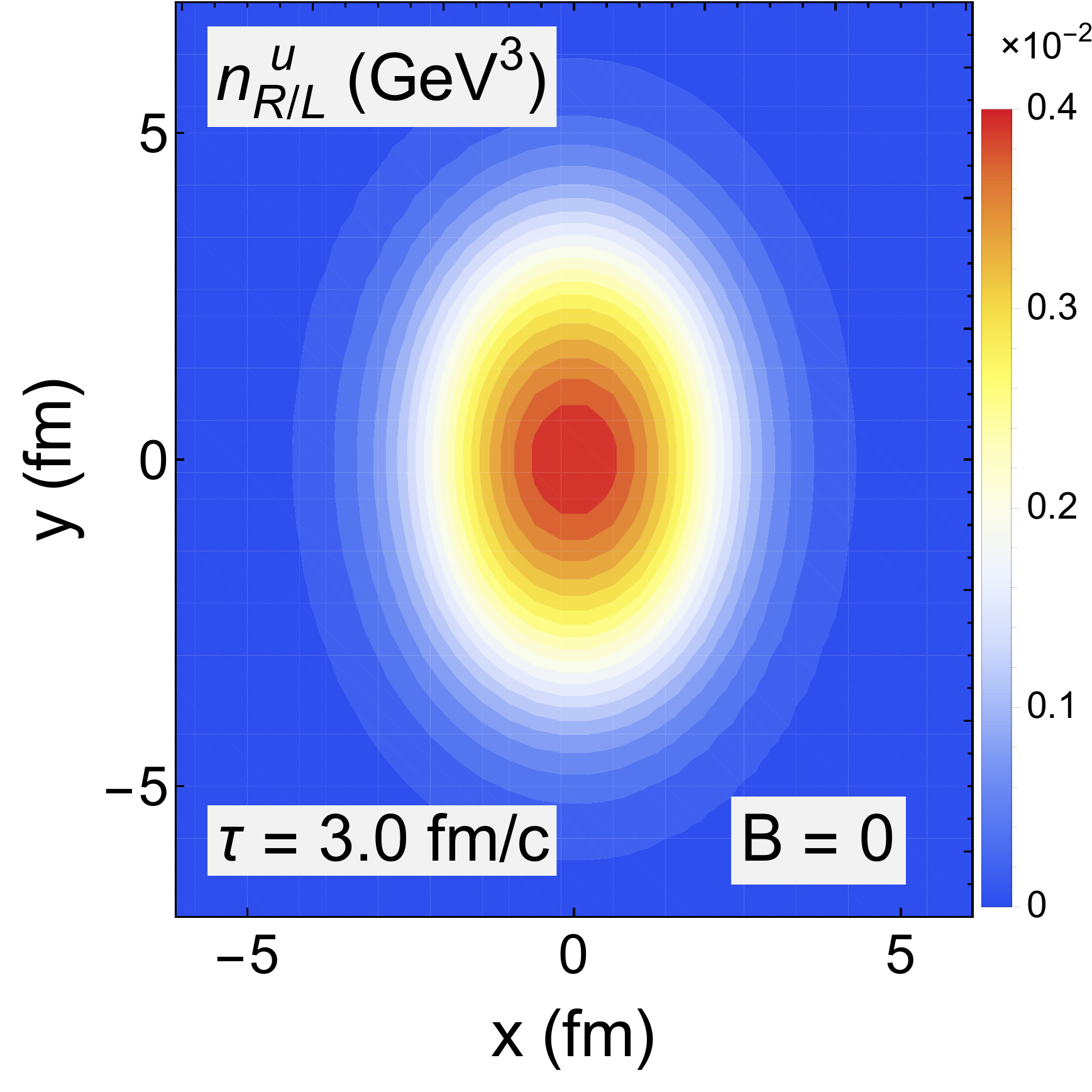}
\includegraphics[width=0.23\textwidth]{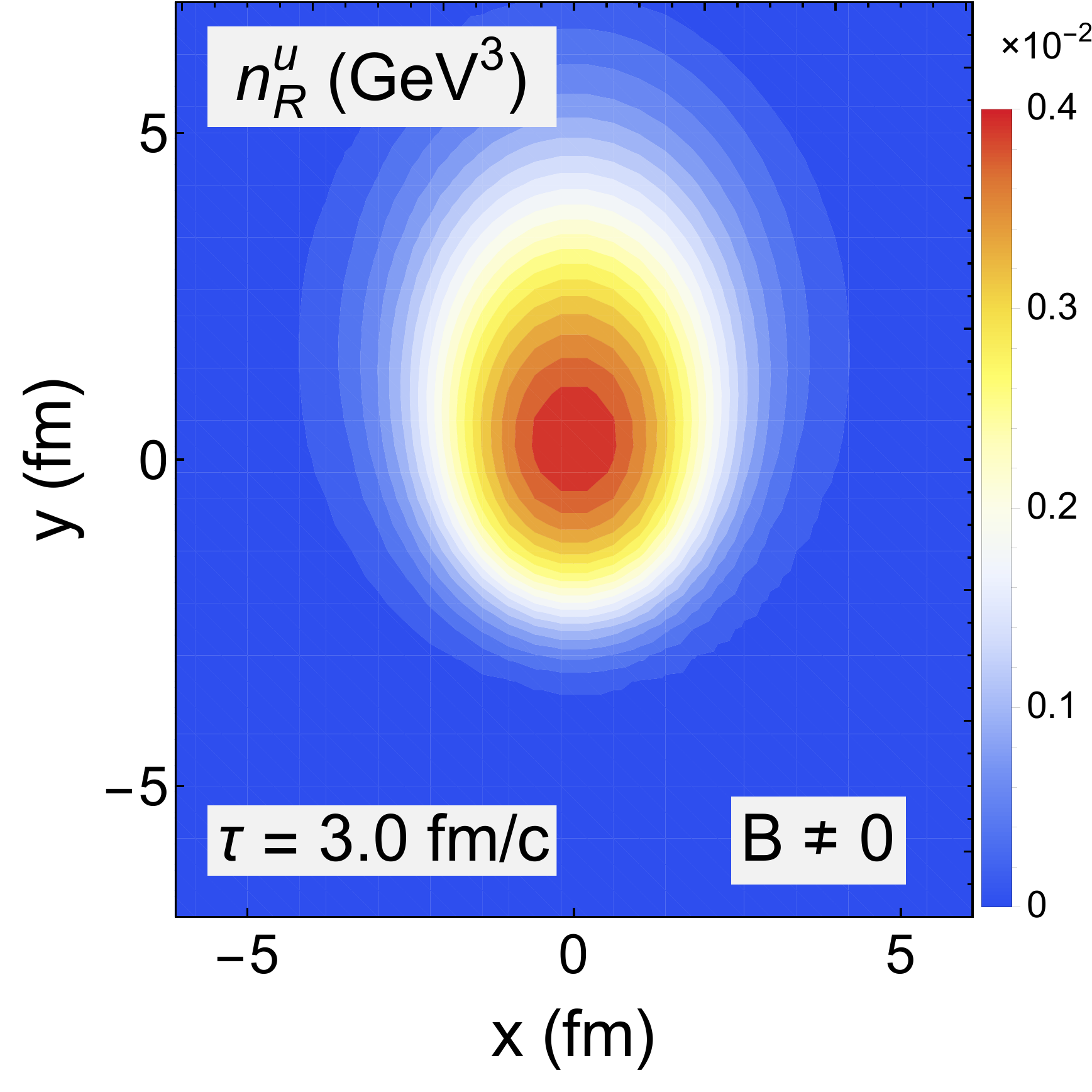} 
\includegraphics[width=0.23\textwidth]{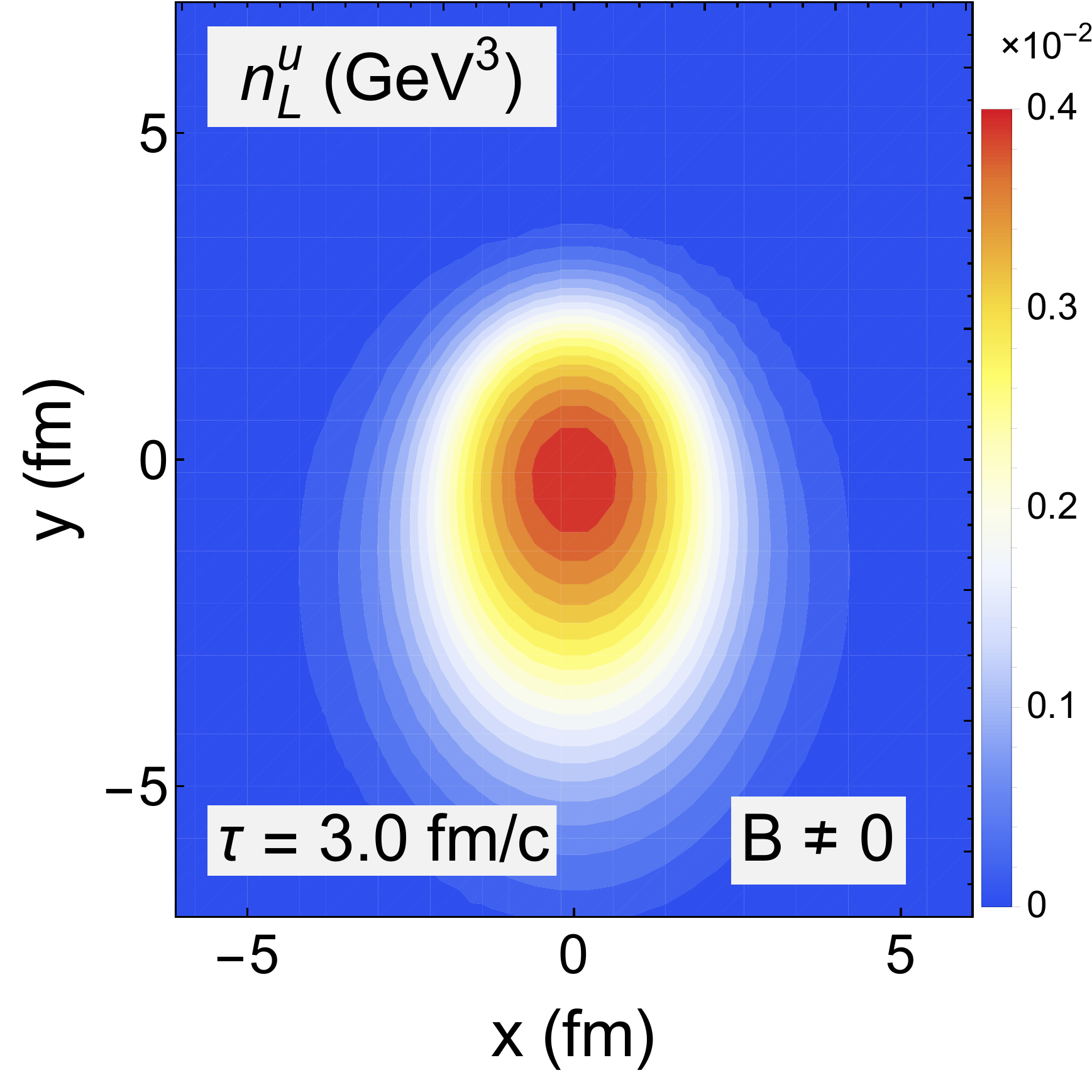}
\caption{(color online) The evolution of $u-$flavor densities via solving AVFD equations from the same initial charge density distribution (for either RH or LH) at $\tau=0.60\rm fm/c$ (left most panel) in three  cases: (a) (second left panel) for either RH or LH density at $\tau=3.00\rm fm/c$ with magnetic field $B\to 0$ i.e. no anomalous transport; (b) (second right panel) for RH density and (c)  (most right panel) for LH density, both at $\tau=3.00\rm fm/c$ with nonzero $B$ field along positive y-axis. } \label{fig_evolution} 
\end{center}
\end{figure*}

To illustrate how the charge separation arises from the CME-induced anomalous transport within the AVFD framework, let us visualize how the fermion  densities evolve under normal and anomalous transport in Fig.~\ref{fig_evolution}. When the hydrodynamic evolution starts (at proper time $\tau=\tau_{0,hydro}=0.6$ fm/c), we initialize the RH/LH $u-$quark number density as a symmetric one (shown in the left most panel). If there is no external magnetic field applied, i.e. only normal transport, both RH and LH $u-$quarks expand with the fluid and also experience viscous transport like diffusion, in a symmetric fashion along $x$-/$y$-direction (shown in the second left panel). On the other hand, once an external magnetic field is turned on along the out-of-plane direction $\hat{\boldsymbol{y}}$, the anomalous CME current propagates RH $u-$quarks toward the direction of $B$ field and LH $u-$quarks toward the opposite direction, leading to an asymmetric pattern of the charge distribution along the out-of-plane direction (shown in the two right panels). 

\begin{figure*}[!hbt]
\begin{center}
\includegraphics[width=0.5\textwidth]{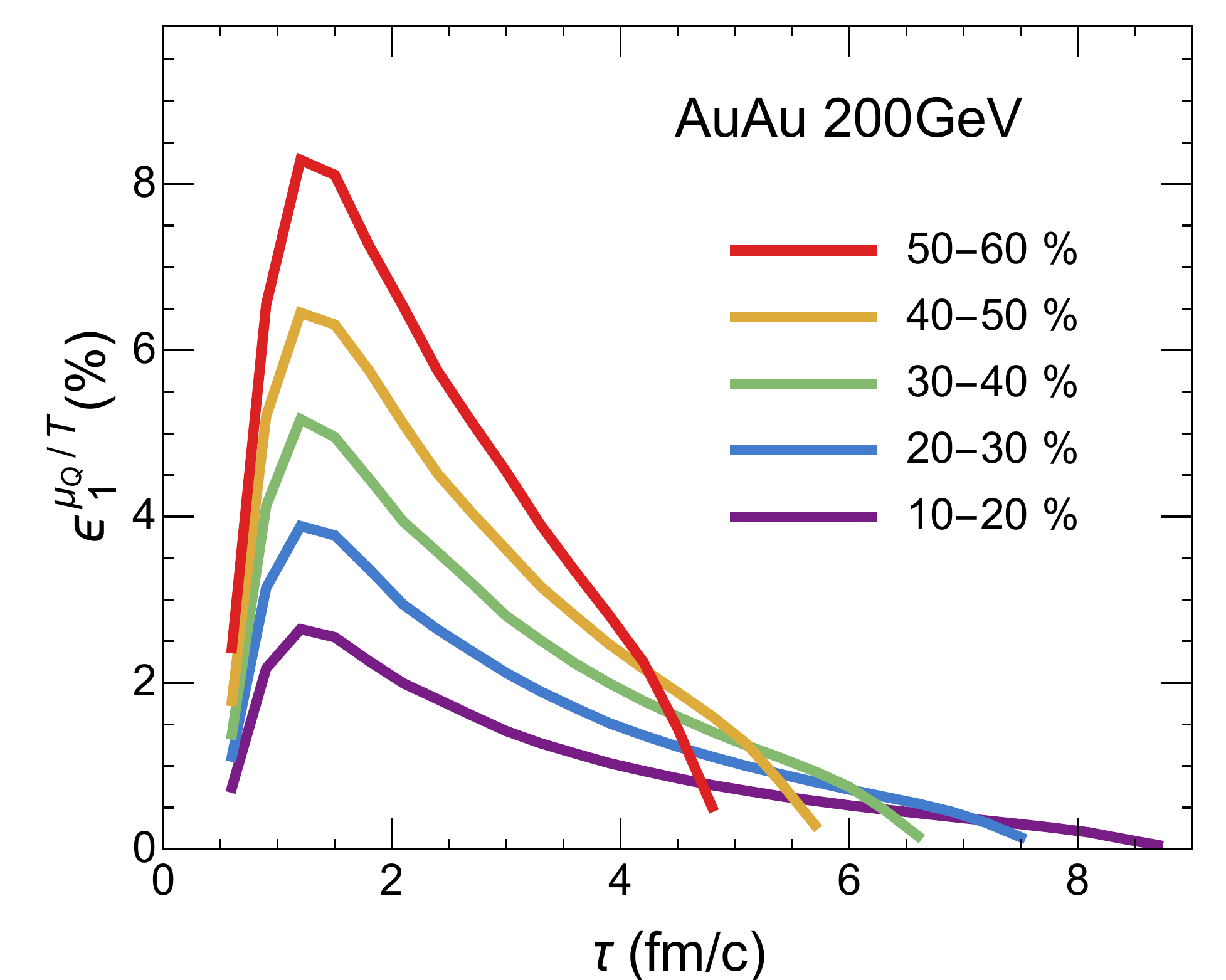}
\caption{(color online) Dipole coefficient of electric chemical potential at freeze-out hyper-surface. Values of magnetic field and chirality imbalance at different centrality range can be found in Sec.~\ref{sec.result.sum}.}
\label{fig_dipole_evolution} 
\end{center}
\end{figure*}

As a result of the anomalous transport under the presence of chirality imbalance (i.e. either the RH or LH pattern in Fig.~\ref{fig_evolution} would dominate), there will be accumulation of opposite charges on the two poles above and below the reaction plane. This would therefore lead to a dipole term in the azimuthal distribution of the electric charge chemical potential, $\mu_{Q} \propto [1+ 2a^{ch}_1 \sin(\phi-\Psi_{RP})]$. 
In Fig.\ref{fig_dipole_evolution} we show the dipole coefficient of the electric chemical potential computed on the freeze-out surface at different proper time $\tau$
\begin{equation}
\epsilon_1^{\mu_Q/T} (\tau) \equiv \frac{1}{2\pi T_{\rm dec}} \int_{T(\tau,\rho,\phi)\equiv T_{\rm dec}} \mu_Q(\tau,\rho,\phi) \sin(\phi-\Psi_{RP})\,\mathrm{d}\phi.
\end{equation}
One can see that such dipole coefficient grows as the accumulation of the CME current. 
At the later stage -- after the magnetic field vanishes -- the electric dipole eventually gets diluted due to diffusion effect as well as expansion of the bulk background.
Upon hadronization via the Cooper-Frye formula~(\ref{eq_cooperfrye}), the dipole term of the chemical potential is converted to the CME-induced charge separation as in Eq.~(\ref{eq_chargedistribution}). 

However, the sign of the CME dipole flips from event to event, according to the sign of chirality imbalance arising from fluctuations in each specific event. Therefore one can only measure this dipole through charge-dependent particle correlations as already discussed before. Nevertheless to quantify the charge separation from CME alone, it suffices to compute the signal with definitive sign of the chirality imbalance i.e. assuming always more RH particles than LH particles with positive $\mu_5$. For events with negative $\mu_5$, the charge separation $a_1$ changes its sign accordingly but the contribution to the two particle correlations would be the same as the events with positive $\mu_5$. 

In passing, let us mention that there have been a number of early attempts in applying the anomalous hydrodynamic framework toward describing charge transport in  heavy ion collisions~\cite{Yin:2015fca,Hirono:2014oda,Yee:2013cya,Hongo:2013cqa}. The AVFD framework developed in \cite{Jiang:2016wve} and in the present paper is by far the most matured approach for state-of-the-art simulations of anomalous transport in heavy ion collisions based on realistic bulk evolution and incorporating normal viscous transport effects simultaneously.  Such a framework allows a quantitative understanding of the generation as well as evolution of CME-induced charge separation signal in the hydro evolution stage as well as its dependence on various ingredients such as the initial conditions, the magnetic fields, as well as the viscous transport coefficients, as will be reported with great details in the rest of this paper.

\section{Quantifying Chiral Magnetic Effect with AVFD}\label{sec.result}

With the above AVFD framework, we are now ready to explore the quantitative aspects of the CME-induced charge separation effect in heavy ion collisions. There are a number of important model inputs: first, the magnetic field strength as well as its time dependence; second, the initial conditions for the fermion charge densities; lastly the two viscous transport parameters namely the diffusion coefficient as well as the relaxation time. In addition there is a hadronic re-scattering stage after the hydrodynamic freeze-out, the influence of which needs to be understood. In this Section, we will systematically investigate the influences of all these factors and discuss the corresponding theoretical uncertainties. Finally, based on our best choices for such model input, we will quantitatively compute the CME-induced H-correlator and compare the results with available experimental data. 

\subsection{Influence of the Magnetic Field}

As the ``driving force'' of the CME current, the strength and space-time dependence of the magnetic field are among the most crucial factors in quantifying  the CME signal. In a heavy ion collision the two colliding nuclei are positively charged and move at nearly the speed of light, thus producing extremely large magnetic fields in the collision zone. For example, the peak value of such $\vec{\bf B}$ field reaches as high as $eB\sim5m_\pi^2$ (or $B\sim10^{15}$ Tesla) at $\sqrt{s_{NN}}=200$ GeV collisions at RHIC, and $\sim 70m_\pi^2$ at $\sqrt{s_{NN}}=2.76$TeV LHC collisions.

\begin{figure}[!hbt]
\begin{center}
\includegraphics[width=0.32\textwidth]{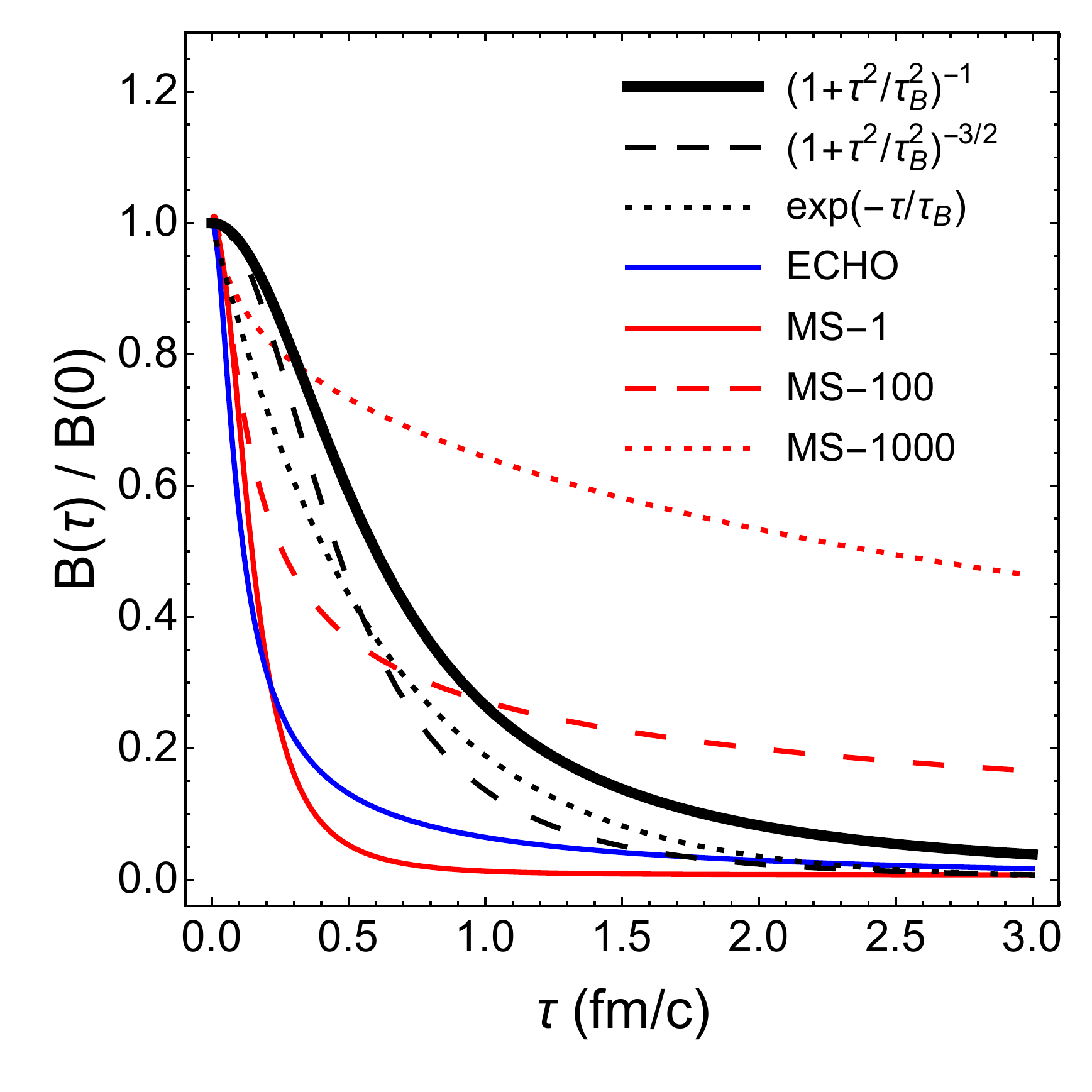}
\includegraphics[width=0.32\textwidth]{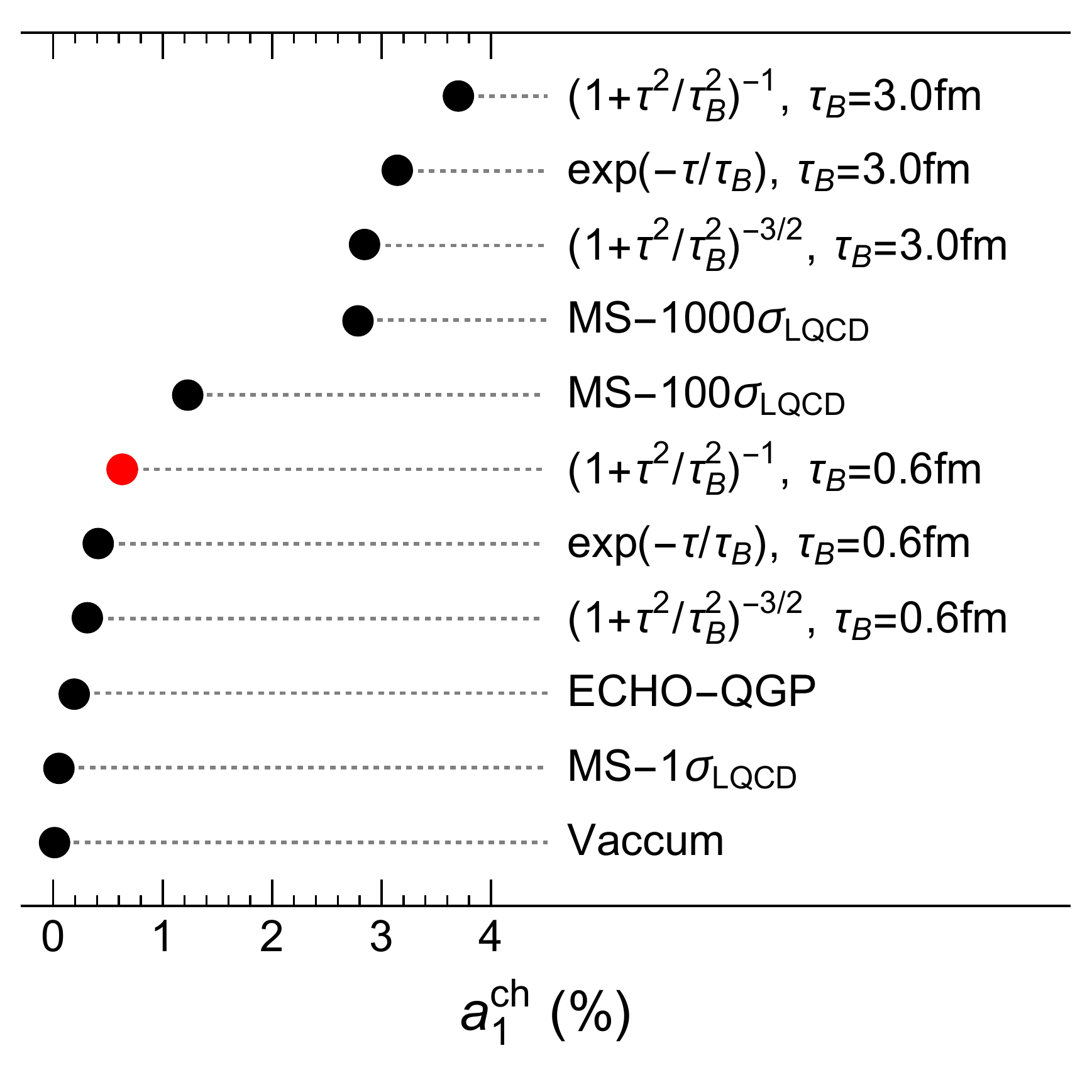}
\includegraphics[width=0.32\textwidth]{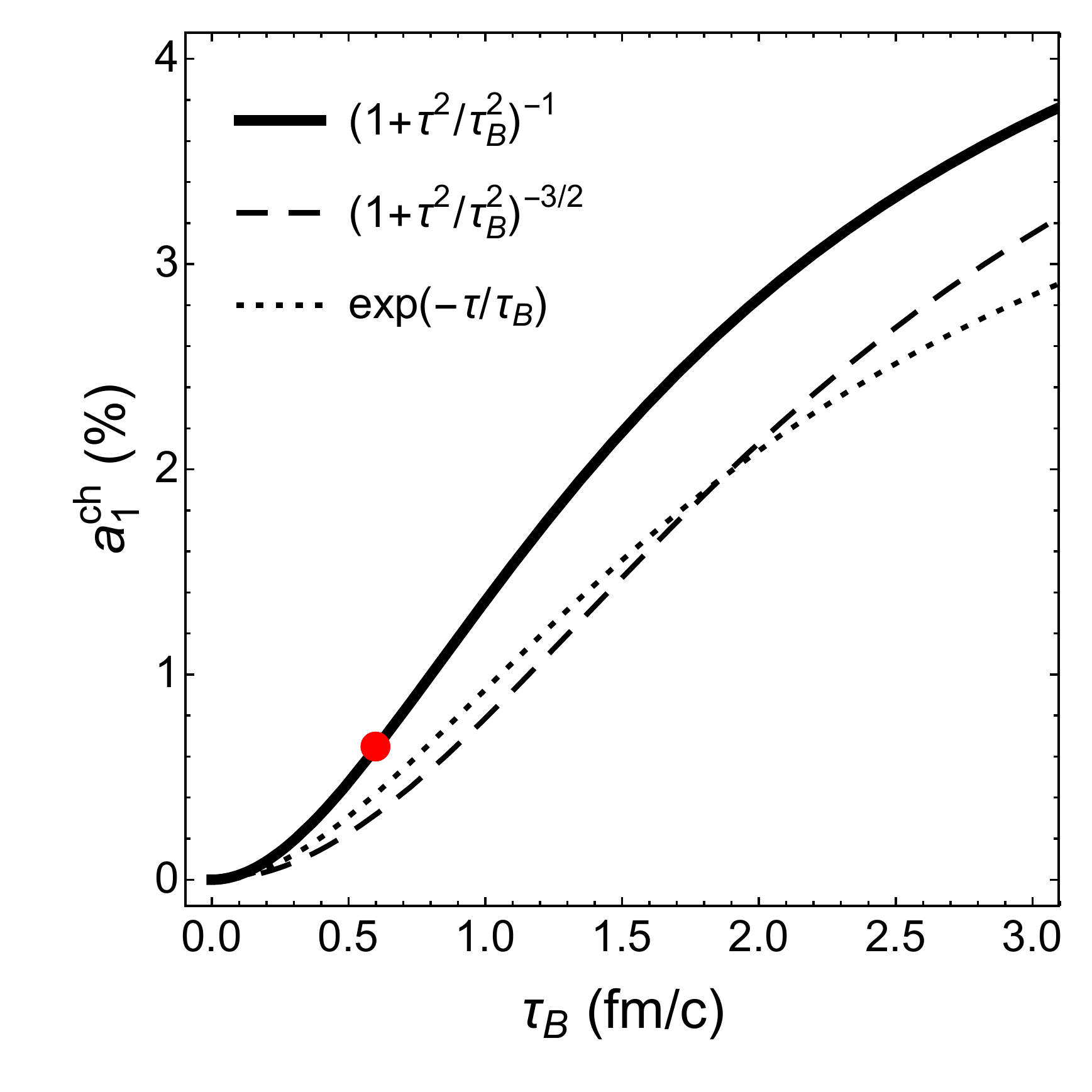}
\caption{(color online) \label{fig_taub}
(left) The time dependence of the magnetic field from different study: ECHO-QGP \cite{Inghirami:2016iru} (blue curve), 
McLerran-Skokov \cite{McLerran:2013hla} with different electric conductivity (solid red curve for $\sigma=\sigma_\text{LQCD}$, dashed red line for $\sigma=100\sigma_\text{LQCD}$, and dotted red line for $\sigma=1000\sigma_\text{LQCD}$.) Also, the thick, dashed, and dotted black curves represent the formulations $B \propto (1+\tau/\tau_B)^{-1}$, $(1+\tau^2/\tau_B^2)^{-3/2}$, and $\exp(-\tau^2/\tau_B^2)$ respectively, with $\tau_B=0.6$ fm/c.\newline 
(middle)  Comparison of charge separation predicted by different time dependences of magnetic field. \newline
(right) Charge separation $a_1$ predicted by different magnetic field life time $\tau_B$. The thick, dashed, and dotted black curves correspond to the formulations $B \propto (1+\tau^2/\tau_B^2)^{-1}$, $(1+\tau^2/\tau_B^2)^{-3/2}$, and $\exp(-\tau/\tau_B)$.  \newline
In both middle and right panels, the red dots correspond to the magnetic field time dependence to be used for the rest of this study.
}
\end{center}
\end{figure}

Many calculations have been done to quantify the magnetic field. For example, by using  event-by-event simulations with Mont-Carlo Glauber model, the  peak value of the magnetic field is well determined, and its azimuthal orientation with respect to event-wise bulk geometry has been quantified to be roughly along out-of-plane direction with a de-correlation factor~\cite{Bloczynski:2012en}. Both its magnitude and its direction are also found to vary only very mildly in the collision overlapping zone except near the fireball edge.   Nevertheless, the $\vec{\bf B}$ time evolution remains  an open question. The main source of the magnetic field, namely the spectator nucleons, pass through each other and fly away quickly from the collision zone at mid rapidity. As a result, the field strength from  such external sources decays rapidly with a behavior roughly following  the formula $B(\tau)=B(0)/(1+\tau/\tau_B)^{3/2}$, with $\tau_B\sim R_\text{nuclei}/\gamma$. On the other hand, the hot medium created in the  collision is a conducting plasma and therefore could in principle delay the decrease of the magnetic field through the generation of an induction current in response to the changing magnetic field. To fully address this issue, one needs to treat both the medium and the magnetic field as dynamically evolving together. 
While many efforts have been made to compute the time dependence of the magnetic field~\cite{McLerran:2013hla,Gursoy:2014aka,Tuchin:2015oka,Inghirami:2016iru}, the answers from different studies vary considerably.        
To get an idea of the current status, we show in Fig.~\ref{fig_taub} (left panel) a comparison of various  results for the time dependence of the magnetic field: the study by McLerran-Skokov \cite{McLerran:2013hla} with conductivity $\sigma=\sigma_{LQCD}, 10^2\sigma_{LQCD}$, and $10^3\sigma_{LQCD}$, and ECHO-QGP simulation \cite{Inghirami:2016iru}, as well as  three types of parameterizations $B \propto (1+\tau^2/\tau_B^2)^{-1}$, $(1+\tau^2/\tau_B^2)^{-3/2}$, and $\exp(-\tau/\tau_B)$. Clearly, the stronger the medium feedback is, the longer the magnetic field lasts. 

The lifetime of the magnetic field strength has a direct and significant impact on the anomalous transport and thus the CME-induced charge separation signal in the end. The AVFD tool allows a quantitative calibration on  influence of the uncertainty in magnetic field lifetime on the predicted  CME signal. In Fig.~\ref{fig_taub} (middle panel), we compare the results for charge separation $a^{ch}_1$ computed with the various different choices of the magnetic field time dependence. Note all these calculations are done with the same initial axial charge condition $n_5/s=0.1$ and with the same peak value of $\vec{\bf B}$ field at time $\tau=0$. 
As a consequence of the huge difference in the $\vec{\bf B}$ field ``surviving'' time, the obtained charge separation signal  varies substantially across various time dependence schemes.  
For the three different $\vec{\bf B}$ parameterizations we further compare them, in Fig.~\ref{fig_taub} (right panel), by showing the  $a^{ch}_1$ versus the magnetic field lifetime parameter $\tau_B$.  Clearly the CME signal grows rapidly with $\tau_B$ in all three cases while for   the same  life-time $\tau_B$  these three parameterizations still show visible difference.

At present there is no convincing conclusion yet regarding the exact time dependence (perhaps except those over-optimistic scenarios), but the comparative study provides a good idea of the associated uncertainties. In the rest part of this paper, we will use a relatively intermediate case  of the following parameterization:  
\begin{equation}\label{eq_taub}
\boldsymbol{B} = \frac{B_0}{1+\tau^2/\tau_B^2} \boldsymbol{\hat{y}}
\end{equation} 
and we adopt a rather ``conservative'' choice  that the life-time of the magnetic field is comparable to the starting time of the hydrodynamic evolution $\tau_{0,hydro}$, i.e. with $\tau_B=\tau_{0,hydro}=0.6$ fm/c (as represented by the red dots in Fig.~\ref{fig_taub} (middle and right panels).  The   peak value of the magnetic field $B_0$ for each centrality is taken from the  event-by-event Monte-Carlo simulations where the field value is already  projected along the elliptic event-plane to account for the angular de-correlation between field direction and bulk geometry due to fluctuations~\cite{Bloczynski:2012en}.

\subsection{Influence of the Initial Conditions}

Another key ingredient of the CME-induced transport  is the chirality imbalance, or more generally speaking, the initial conditions of RH/LH charge densities (or equivalently the initial conditions for the vector and axial charges of each flavor of fermions). 
In this section, we study the influence of initial conditions of charge distribution on the final hadron charge separation with the AVFD tool.  For this framework, one needs to provide the following initial conditions, namely the initial four-current $J^\mu_{\chi,f}(\tau=\tau_0)$ for each flavor $f$ as well as chirality $\chi$. For most part of this study, we use null initial condition for the spatial three-current components i.e. setting all $\vec{\bf J}_{\chi,f}(\tau_0)\to 0$ (except in Sec.~\ref{sec.prehydro}) while only consider the zeroth component i.e. the number densities $J^0_{\chi,f}(\tau=\tau_0)$. Equivalently one can set for each flavor the initial vector and axial charge densities $n=J^0_R+J^0_L$ and $n_5=J^0_R-J^0_L$. In the following we discuss the influence of the vector and axial charge initial conditions respectively.

\subsubsection{Vector Charge Initial Conditions}

\begin{figure}[!hbt]
\begin{center} 
\includegraphics[width=0.5\textwidth]{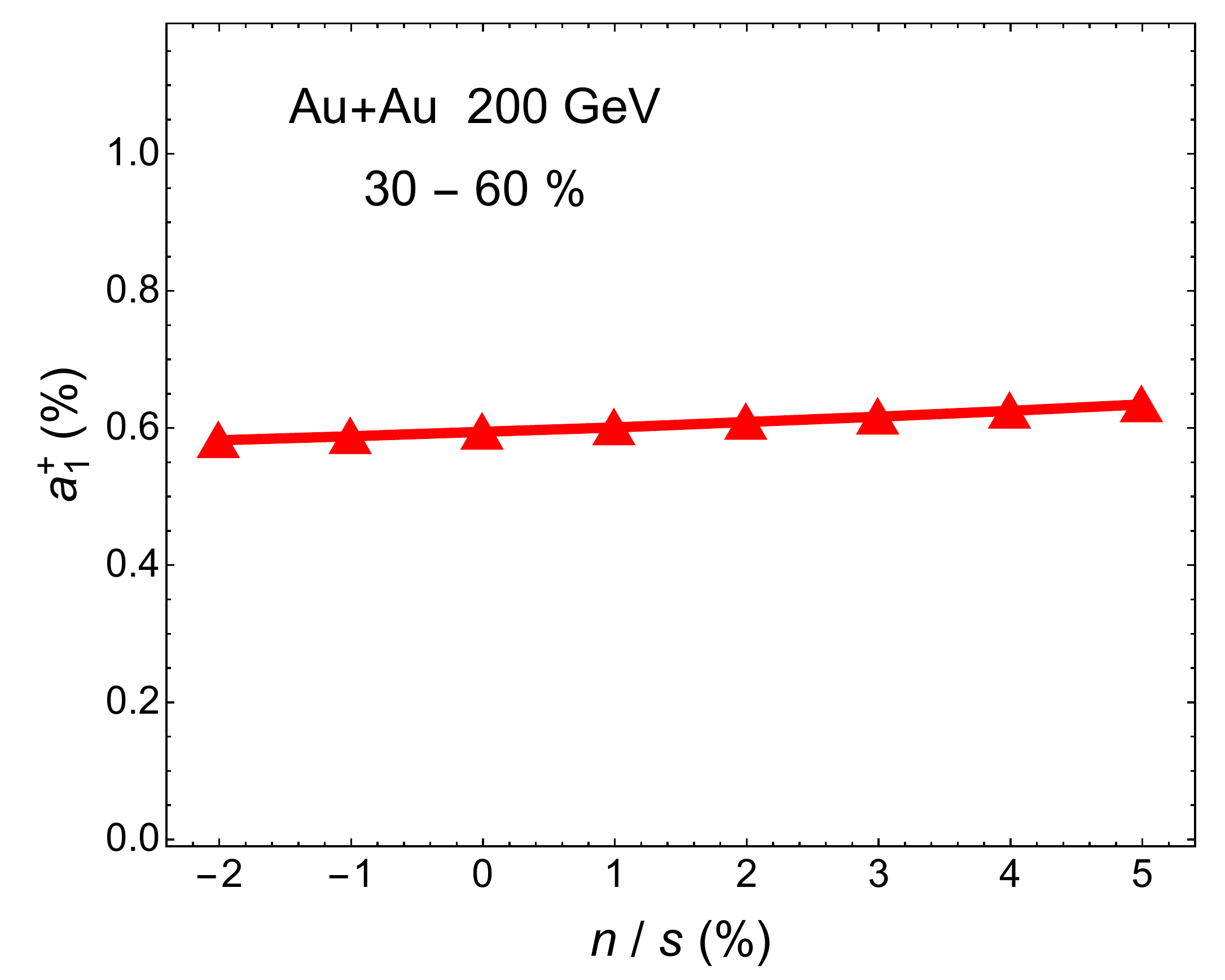}
\caption{(color online) Charge separation $a^{ch}_1$ versus initial vector charge density per flavor.}  
\label{fig_nv}
\end{center}
\end{figure}

As demonstrated earlier in this paper, it is the (vector) chemical potential at the freeze-out hyper-surface that directly affects the observed  particle yields as well as the   CME-induced charge separation $a_1$. Such chemical potential is determined from the corresponding vector number density, the number density of all species of quarks including both RH and LH sectors. One would be interested in how the initial conditions of vector charge affects the CME signal. 
As a test, we take the initial condition for the vector charge densities to be proportional to the entropy density at the hydro initial time, and then vary this proportionality coefficient to examine its effect on the final charge separation $a^{ch}_1$.  
In Fig.~\ref{fig_nv} we show the $a^{ch}_1$ computed from AVFD with a wide range of different initial vector charge densities (but with the same fixed initial axial charge density $n_5/s=0.1$). The resulting signal stays roughly constant despite even significant changes of the vector densities. Clearly this suggests an insensitivity of the CME signal to the initial vector charge densities, which would therefore imply a negligible potential influence from the uncertainty in constraining the initial conditions for vector charge densities.  For most AVFD simulations performed in this study, we set a small but nonzero initial vector charge density as $n_{u,d}/s = 1\%$ (due to stopping) which is a very reasonable estimate for top energy collisions at RHIC as also indicated by e.g.   AMPT simulations.

\subsubsection{Axial Charge Initial Conditions}

The axial charge quantifies the number difference between RH and LH fermions. As discussed before, it plays a key role in the generation of the CME current and hence the final charge separation signal which is expected to sensitively depend upon the initial condition for the axial charge. 

\begin{figure}[!hbt]
\begin{center}
\includegraphics[width=0.5\textwidth]{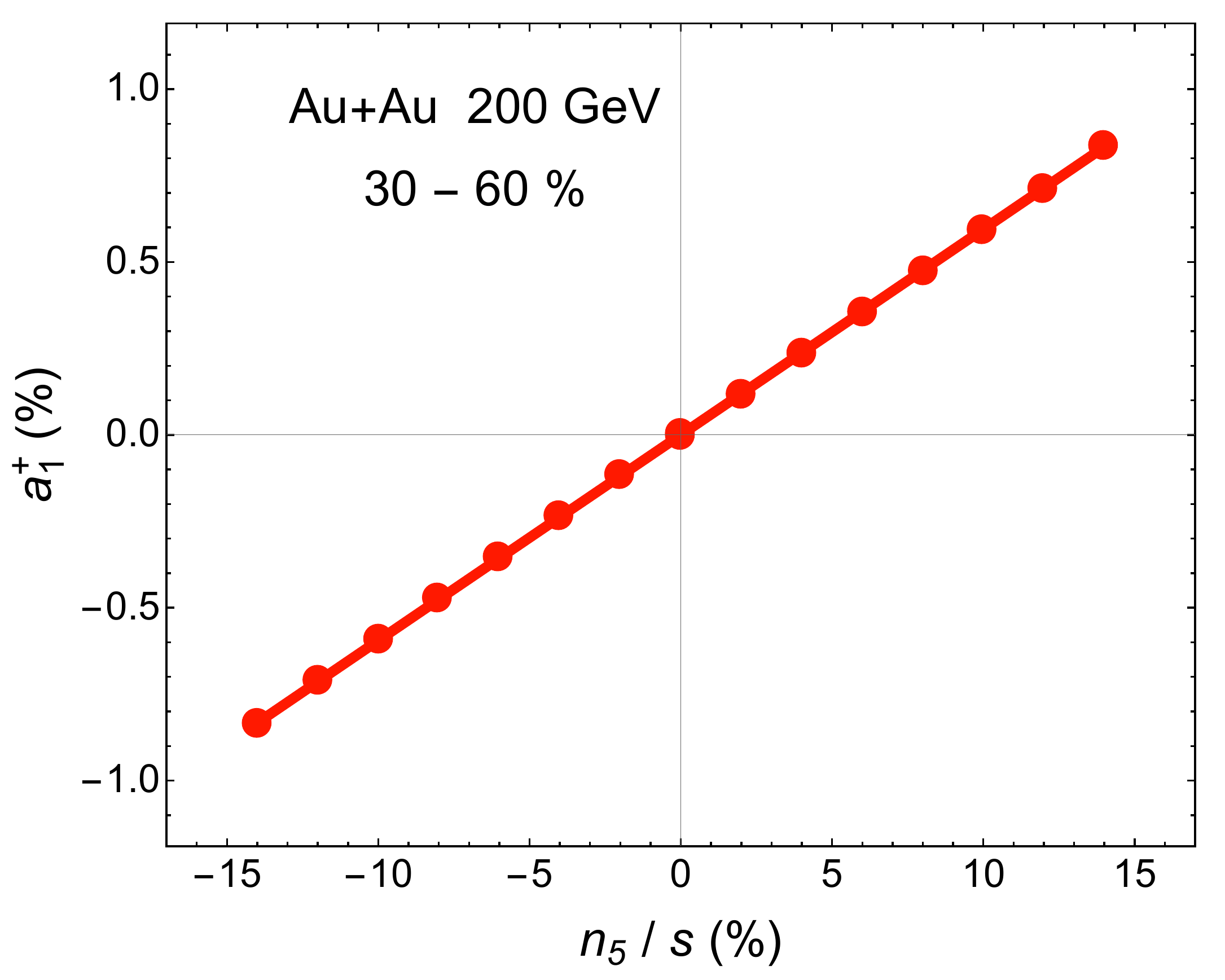}  
\caption{(color online) Charge separation $a_1$ versus initial axial charge density per flavor.}
\label{fig_na}
\end{center}
\end{figure}

For what we consider, the axial charge density is small compared with entropy density (or equivalently the corresponding chemical potential being small compared with temperature), and therefore the axial charge density and axial chemical potential are linearly proportional to each other. Thus one expects $J^\mu_\text{CME} = C_A \mu_5 B^\mu \propto n_5$, i.e. the final charge separation signal should be roughly linearly dependent on the amount of initial axial charge.  This linear dependence is indeed verified to be true, as shown in   Fig.~\ref{fig_na}. When $n_5=0$, there is no chiral imbalance, and no CME-induced charge separation as expected.

As demonstrated above, the CME charge separation is mainly controlled by chirality imbalance  while insensitive to the vector charge initial condition, so let us focus on how to properly estimate the axial charge initial condition. Following the scenario based on chirality imbalance arising from gluonic topological charge fluctuations in the early-stage glasma~\cite{Kharzeev:2001ev,Muller:2010jd,Hirono:2014oda}, the starting point is the following estimation of the axial charge density fluctuations:
\begin{equation}
\label{eq8}
\sqrt{n^{2}_{5}}
=\left[ \tau_{0}\, \frac{\, (g E^{c})\, (g B^{c})\,}{16\pi^{2}}\right]  \times \sqrt{N_{\rm tube}} \times \frac{\pi\, \rho^{2}_{\rm tube}}{A_{\rm overlap}}\, . 
\end{equation}
The part inside $\left[\ldots\right]$ counts the density of axial charge $n_{5}$ in a {\em single} glasma flux tube in which the parallel/anti-parallel chromo-electric and chromo-magnetic fields $E^{c}, B^{c}$ provide nonzero topological charge density thus nonzero induce corresponding axial charge density via the standard anomaly formula. The total number of glasma flux tubes $N_{\rm tube}$ in a given event can be estimated from the binary collision number, $N_{\rm tube}\simeq N_{\rm coll}$. Note that the color fields inside glasma flux tube are very strong, i.e. $E^{c}, B^{c} \sim Q_{s}^2/g$.  
However inside each flux tube the $E^{c}, B^{c}$ fields randomly take parallel or anti-parallel configurations and thus gives randomly positive or negative contributions to axial charge: this reduces the net axial charge fluctuation one could get in each event, and such reduction effect is taken in account by the $\sqrt{N_{\rm tube}}$.
Finally, to get an averaged/smeared-out axial charge density over the whole fireball transverse area by the contribution of individual flux tube, we include a ``dilution'' factor $\frac{\pi\, \rho^{2}_{\rm tube}}{A_{\rm overlap}}$. Here $\pi\, \rho^{2}_{\rm tube}$ is for the flux tube transverse area (with $\rho_{tube}\simeq 1 \rm fm$  the transverse extension of a glasma flux tube) while    $A_{overlap}$  is the transverse geometric overlapping area for the collision zone of the two colliding nuclei. Putting these all together, one then obtains the following estimate:
\begin{eqnarray} \label{eq_n5}
\sqrt{ \left< n_5^2 \right> } \simeq  \frac{Q_s^4\, (\pi \rho_{tube}^2 \tau_0) \, \sqrt{N_{coll.}}} {16\pi^2 \, A_{overlap}} \,\, .
\end{eqnarray}  
 The above estimate is then used to determine a ratio $\lambda_5$ of total average axial charge over the total entropy in the fireball at initial time $\tau_0$, $\lambda_5 \equiv \frac{\int_V \, \sqrt{ \left< n_5^2 \right> }}{\int_V \, s}$ where the integration is over fireball spatial volume and the $s$ is the entropy density from bulk hydro initial condition. This ratio is then used in the AVFD simulations to set an   initial axial charge distribution locally proportional to entropy density via  $n_5^{initial} = \lambda_5 \, s$. This properly reflects the fact that axial charge arises from local domains with gluon topological fluctuations and that there are more such domains where the matter is denser.  Such axial charge density estimate depends most sensitively upon the saturation scale $Q_s$, in the reasonable range of  $Q_s^2\simeq 1\sim 1.5 \rm GeV^2$ for RHIC 200GeV collisions~\cite{Rezaeian:2012ji,Kowalski:2007rw}.

The axial charge estimated from Eq.(8), while not large,  is not very small. This is likely due to the out-of-equilibrium nature of the glasma with very strong color fields and large topological fluctuations as compared with usual expectation from perturbative thermal plasma case.  The estimate of initial axial charge used in  Ref.~\cite{Hirono:2014oda}  and here would  correspond to a relatively large Chern-Simons diffusion rate $\Gamma_{\rm CS} \sim Q^{4}_{s} $ which is much larger than the usual thermal values.  Interestingly this rapid rate has been confirmed   recently in Ref.~\cite{Mace:2016svc} which extracts the rate by   using classical-statistical real time lattice simulation in non-equilibrium Glasma of weakly coupled but highly occupied gauge fields.

\subsection{Dependence on the Viscous Transport Parameters}
 
While the main new interest and recent developments focus on the anomalous transport and its consequence on the charge distribution in QGP, it is quite obvious that more conventional viscous transport like charge diffusion would certainly also affect the charge distribution.  However such contributions are largely ignored in past studies and it was unclear to which extent the CME signal would depend on normal viscous transport. The AVFD framework for the first time allows quantitative study of this problem and helps constraining the important theoretical uncertainty due to the viscous transport. 
 
As shown in Eq.(\ref{eq_IShydro_0}-\ref{eq_avfd_ns_2}), the AVFD framework is based on second-order Isreal-Stewart \& Navier-Stokes equations. Hence it includes the diffusion and conduction effects in the first order as well as the relaxation effect in the second order. Such viscous transport is controlled by two key parameters: the diffusion coefficient and the relaxation time.  In this subsection, we quantify how CME signals are influenced by these parameters.

\subsubsection{Dependence on the diffusion coefficient}

Diffusion effect is the macroscopic manifestation of the Brownian motion of the particles. It causes the conserved charge density to spreads out under the presence of density gradient, leading eventually to homogeneous distribution in thermal equilibrium. The diffusion coefficient $\sigma$ controls how fast the diffusion process transports charges around.  

\begin{figure}[!hbt]
\begin{center} \label{fig_sig}
\includegraphics[width=0.5\textwidth]{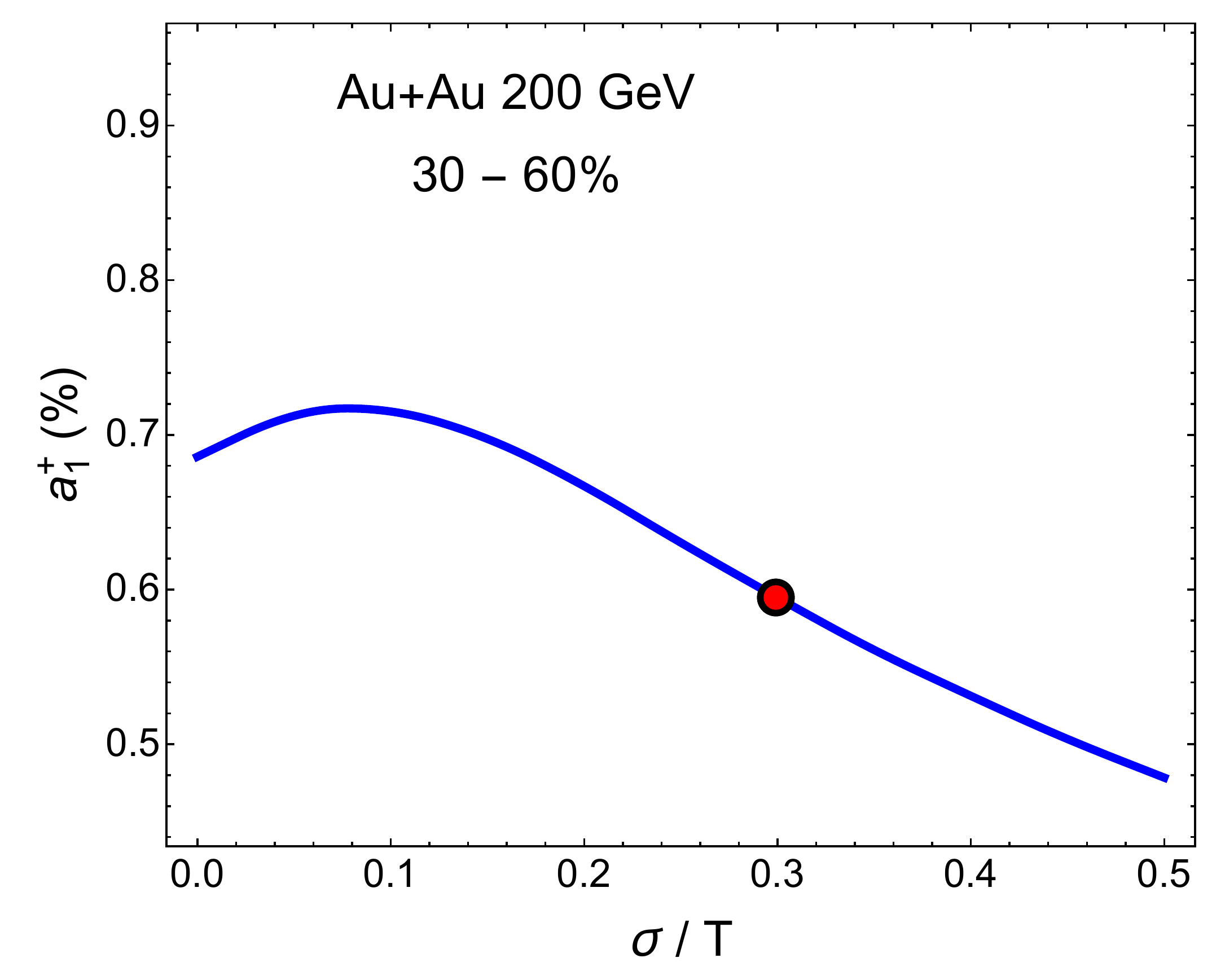}  
\caption{(color online) The charge separation $a_1$ as a function of the diffusion coefficient $\sigma$ (scaled by the temperature $T$). The red dot indicates the commonly chosen value, to be used later in the paper. }  \label{fig_diffusion}
\end{center}
\end{figure}

The  dependence of the charge separation $a_1$ on the diffusion coefficient $\sigma$, is shown in Fig.~\ref{fig_diffusion}. A large value of $\sigma$ would imply strong and fast diffusion for any nonzero charge density and therefore would suppress the charge separation induced by CME by transporting net charges across the reaction plane. Indeed we see that the signal $a_1$ decreases with increasing $\sigma$ when the diffusion effect is strong. On the other hand, when $\sigma$ is small, the diffusion is not strong enough to bring net charges from one side of the reaction plane to the other side. Instead it can help the  CME-induced charge dipole spread out a little more over the freeze-out hyper-surface and slightly enhance the $a_1$ of the final hadrons.  One can see that such small diffusion effect indeed slightly increases the charge separation signal.

\subsubsection{Dependence on the relaxation time}

\begin{figure}[!hbt]
\begin{center} \label{fig_tau}
\includegraphics[width=0.5\textwidth]{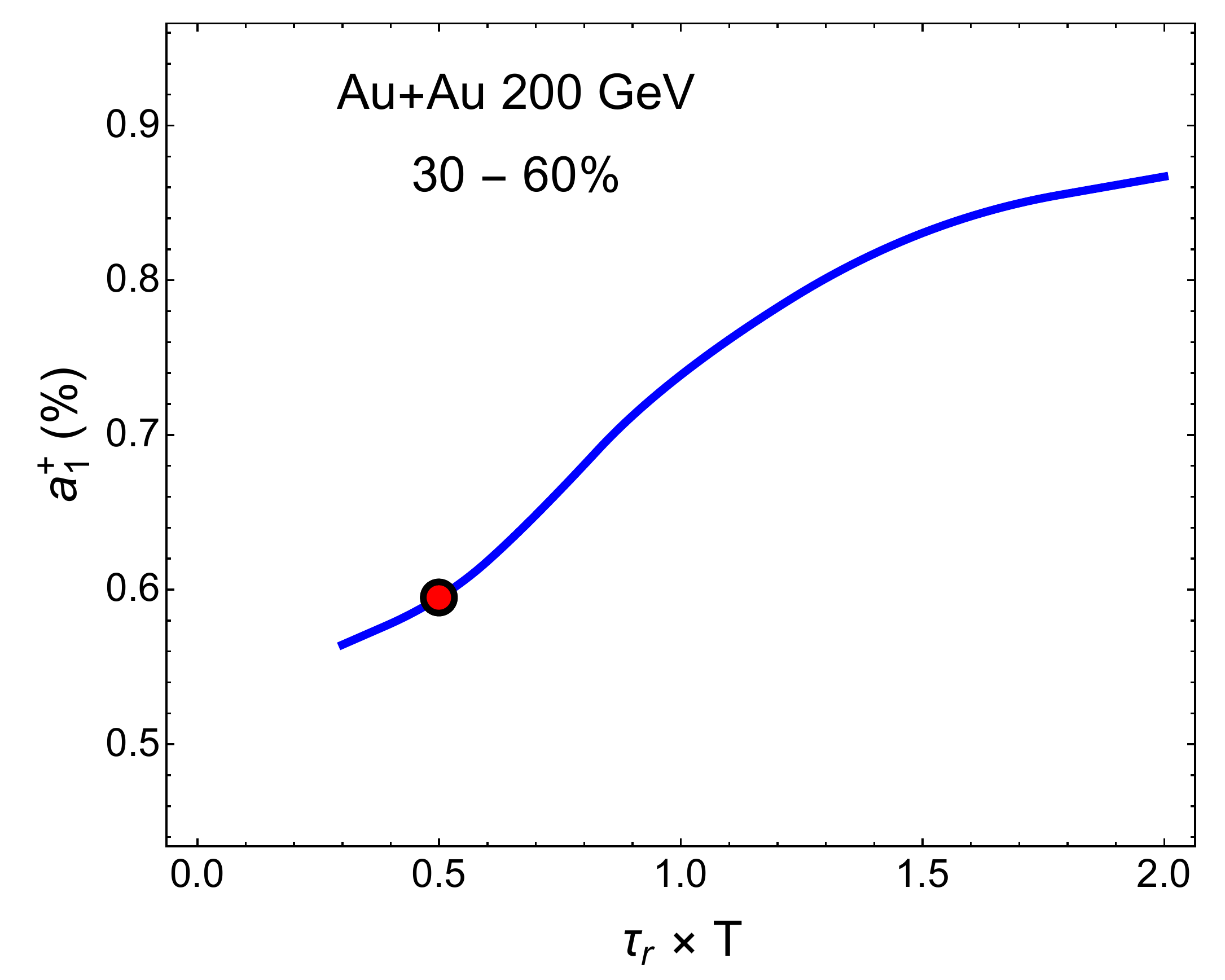}  
\caption{(color online) The charge separation $a_1$ as a function of the relaxation time $\tau_r$ (scaled by $T^{-1}$). The red dot indicates the commonly chosen value, to be used later in the paper.}  \label{fig_relaxation}
\end{center}
\end{figure}

Relaxation time parameter $\tau_r$ controls the   time scale that is needed to build up the ``diffusion current'' in response to the density gradient. Intuitively speaking, a small (large) $\tau_r$ implies rapid (slow) building up of the diffusion current and causes stronger (weaker)  diffusion effect, thus suppressing more (less) the CME-induced charge separation. In deed, as shown in Fig.~\ref{fig_relaxation}, the signal $a_1$ increases steadily with increasing relaxation time.

In summary, one can see that both the diffusion and relaxation effects have considerable influence on the charge separation signal.
A commonly adopted choice of $\sigma=0.3T$ and $\tau_r=0.5/T$ (see e.g. \cite{Denicol:2012vq}), as indicated by the red blob in Fig~\ref{fig_diffusion}\&\ref{fig_relaxation}, will be used in our later computations. It is worth emphasizing that the choice of diffusion parameter here corresponds to the electric conductivity $\sigma_{\rm ele}=e^2\sigma\sim5$ MeV, which is consistent with the conductivity given by lattice simulations (see e.g. \cite{Amato:2013naa}). The curves in Fig~\ref{fig_diffusion}\&\ref{fig_relaxation} shall give a quite clear idea of the uncertainty in the signal due to the  uncertainty associated with the input values of $\sigma$ and $\tau_r$.

\subsection{Contribution from resonance decay}

\begin{figure}[!hbt]
\begin{center}
\includegraphics[width=0.4\textwidth]{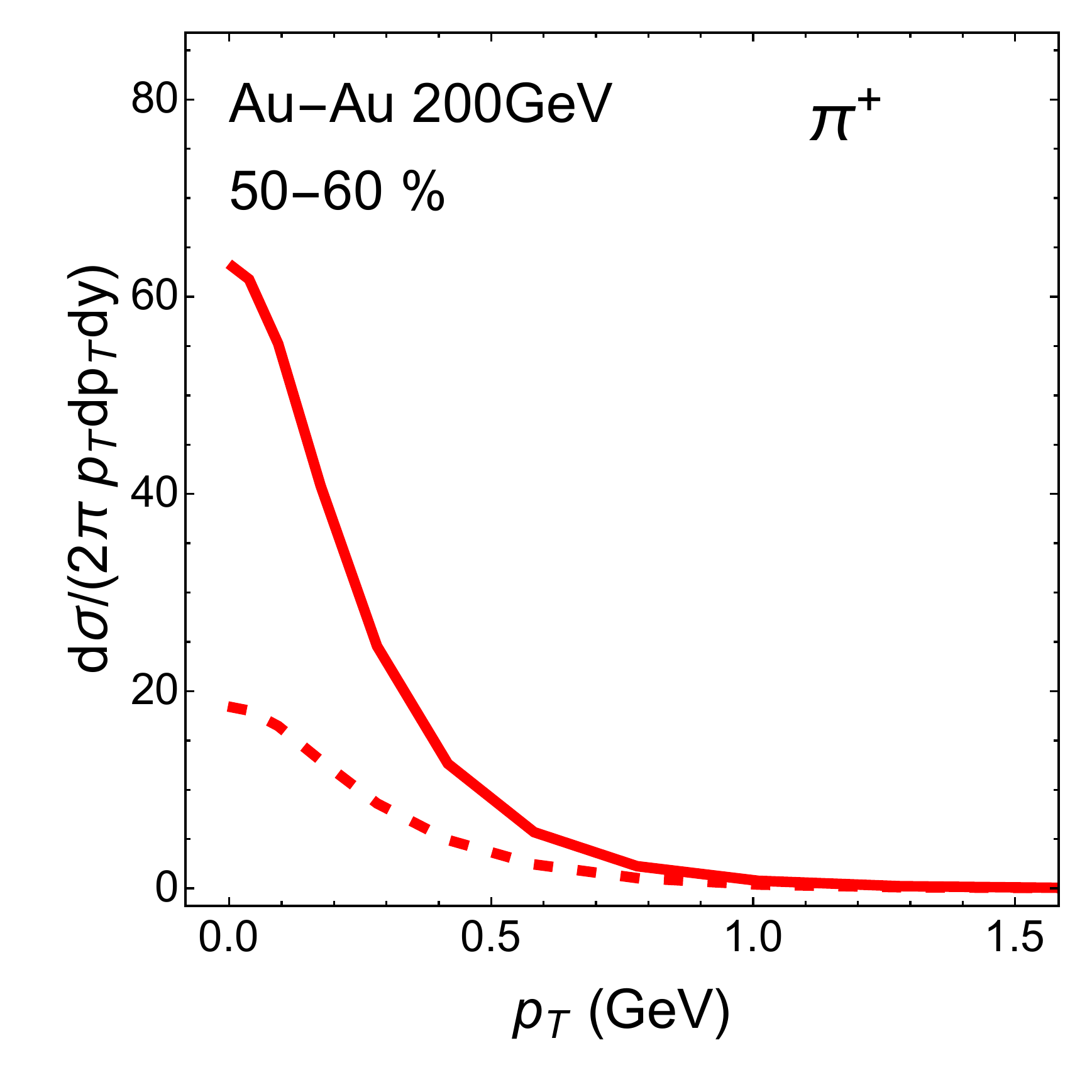}
\includegraphics[width=0.4\textwidth]{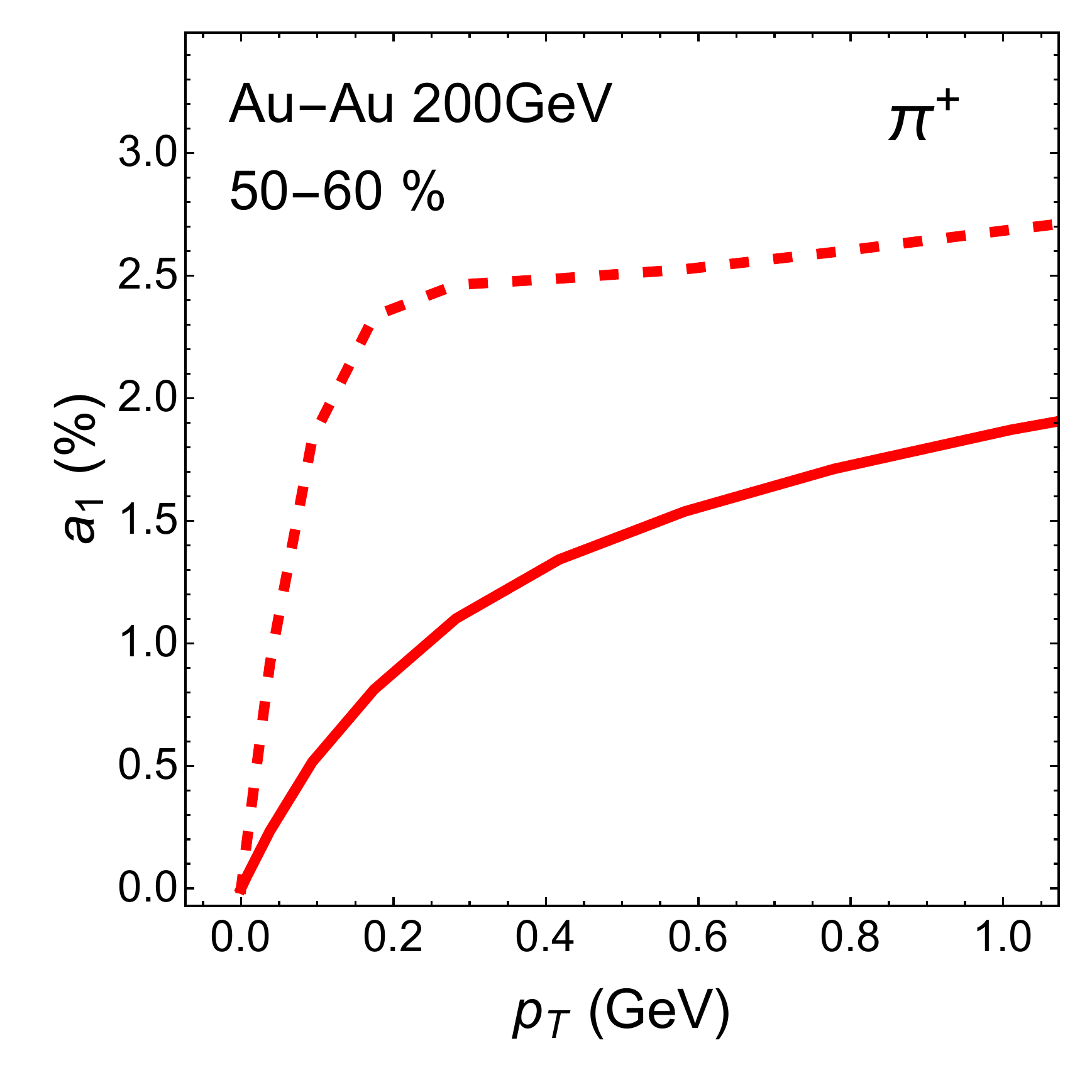}
\caption{(color online) Comparison of differential cross-section (left) and charge separation signal $a_1$ (right) for $\pi^+$ directly produced at the freeze-out surface (dashed curves), versus the final observables   including resonance decay contributions (solid curves).} \label{fig_resonance}
\end{center}
\end{figure}

In additional to the viscous transport, another ``trivial''/conventional effect which was often not included properly in previous simulations but which bears quantitative consequences,  is the contribution from resonance decay in the hadronic cascade stage. 
As the lightest and most abundant particle, the finally observed pions receive a  substantial contribution from the feed-down of resonance decays (see Fig~\ref{fig_resonance}(left)), which clearly affect the various bulk observables (usually dominated by pions) such as  the harmonic flow coefficients as well as the charge distributions.  With the resonance decays already implemented in the VISH2+1 code, the AVFD simulation takes such effect into account and thus allows investigation of their impact on the desired charge separation signal.

In Fig.~\ref{fig_resonance}(right) we show the comparison of charge separation $a_1$ computed from $\pi^+$ directly produced at the freeze-out surface, versus that computed from the final observed particles including resonance feed-down. One can see that after taking such effect into account, $a_1$ is suppressed by $\sim 30\%$. Noting that the main source of  decay-produced $\pi^\pm$'s are from the processes like $\rho\to\pi\pi$, $\eta\to\pi\pi\pi$, etc., we take the $\rho^{0,\pm}$ mesons as an example to explain such suppression effect. First of all, the charged $\rho^\pm$ mesons are affected by CME and carry non-zero charge separation $a_1$ in their distributions. However after their decay  into pions by $\rho^{\pm}\to\pi^0\pi^\pm$, the momentum direction of the parent $\rho$ is not perfectly preserved by the daughter pion, thus smearing out the charge separation initially carried by rho mesons. Secondly, the uncharged $\rho^0$ mesons do not carry any CME charge separation and neither do the daughter particles from $\rho^0\to\pi^+\pi^-$, but these decay pions still contribute to the total number of charged pions thus diluting out the observed charge separation signal. In both   cases, the charge separation $a_1$ will be suppressed due to the smearing and dilution, resulting in a significantly reduced magnitude of the signal. This effect must be quantitatively taken into account for meaningful predictions and comparison with experimental data.

\subsection{Quantifying the CME  signal}\label{sec.result.sum}

Given the above detailed investigations on the various theoretical inputs and how they influence the charge separation at quantitative level, we now proceed to quantify the CME signal with our best constrained parameter choices. For the magnetic field, we assume them to be  homogeneous in space while to evolve  in time according to Eq.(\ref{eq_taub}), with the life-time parameter $\tau_B=\tau_{0,hydro}=0.6$ fm/c. The initial condition for the axial charge is given by Eq.(\ref{eq_n5}), with the saturation scale $Q_s^2$ in a reasonable range of $1\sim 1.5 \rm GeV^2$ for the top RHIC energy collisions. The diffusion coefficient and relaxation time are chosen with  the commonly used values as $\sigma=0.3T$, and $\tau_{r}=0.5/T$. The resonance decay contributions are properly taken into account. Finally,  all the $p_T-$integrated results in this paper (for observables like the  charge separation $a_1$ and two-particle correlation $H$) are computed from hadrons in the range of $0.15<p_T<2{\mathrm GeV}$, which is exactly the same as the experimental kinematic cuts adopted in the relevant STAR measurements.

\begin{figure}[!hbt]
\begin{center} 
\includegraphics[width=0.5\textwidth]{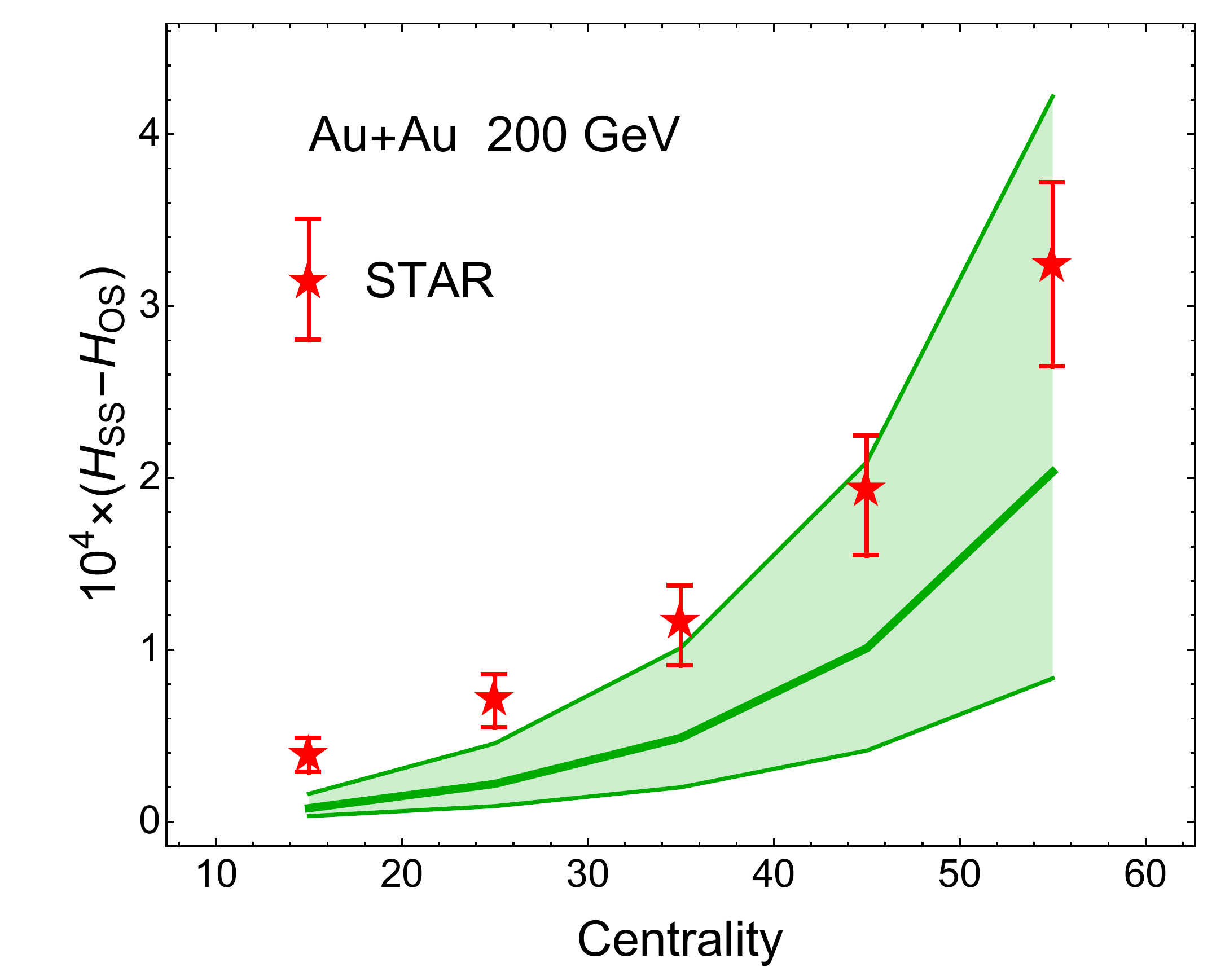}
\caption{\label{fig_H} 
 Quantitative predictions from Anomalous-Viscous Fluid Dynamics simulations for the CME-induced H-correlations, in comparison with STAR measurements~\cite{STAR_LPV_BES}. The uncertainty of experimental data comes from the uncertainty of $\kappa$. Central values correspond to $\kappa=1.2$, while upper and lower bounds are for $\kappa=1$ and $\kappa=1.5$, respectively. The green bands reflect current theoretical uncertainty in the initial axial charge generated by gluonic field fluctuations.}
\end{center}
\end{figure}

\begin{table}[!hbt]\centering
\begin{tabular}{cccccc}
\hline
centrality bin & 10-20\% & 20-30\% & 30-40\% & 40-50\% & 50-60\% \\
\hline
$eB_0(m_\pi^2)$ & 2.34 & 3.10 & 3.62 & 4.01 & 4.19 \\
\hline
$n_5/s$ & 0.065 & 0.078 & 0.095 & 0.119 & 0.155 \\
\hline
\end{tabular} 
\caption{Centrality dependence of magnetic field peak strength and the initial chirality imbalance. The $n_5/s$ shown here is obtained with a  saturation scale $Q_s^2=1.25{\rm GeV}^2$.\label{tab1}}
\end{table}

The AVFD results for the charge dependent H-correlations, obtained with the aforementioned parameters, for various centrality bins are presented in Fig.~\ref{fig_H}, with the green band spanning the range of key parameter $Q_s^2$ in the $1\sim 1.5 \rm GeV^2$ range to reflect the uncertainty in estimating initial axial charge (see Eq.(\ref{eq_n5})). Clearly the CME-induced correlation is very sensitive to the amount of initial axial charge density as controlled by $Q_s^2$, especially in the peripheral collisions. The comparison with STAR data~\cite{STAR_LPV_BES} shows very good agreement for the magnitude and centrality trend, provided that the $Q_s$ value is at the relatively larger end of the quoted band. In Table.~\ref{tab1} we show the values of the magnetic field peak strength and the initial axial charge density (normalized by entropy density) that are used in computing the Fig.~\ref{fig_H}.

\begin{figure*}[!hbt]
\begin{center} 
\includegraphics[width=0.32\textwidth]{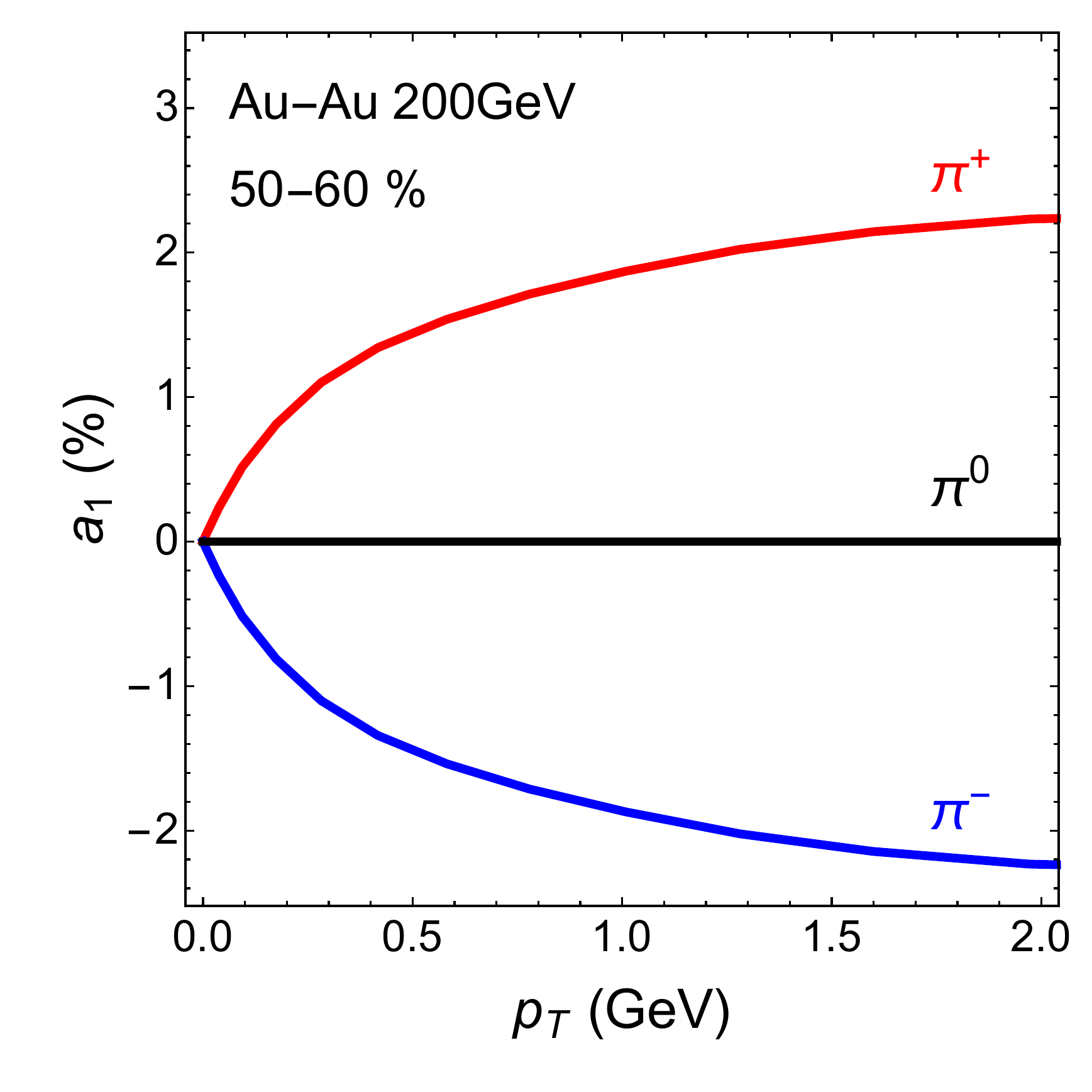} \hspace{0.02in}
\includegraphics[width=0.32\textwidth]{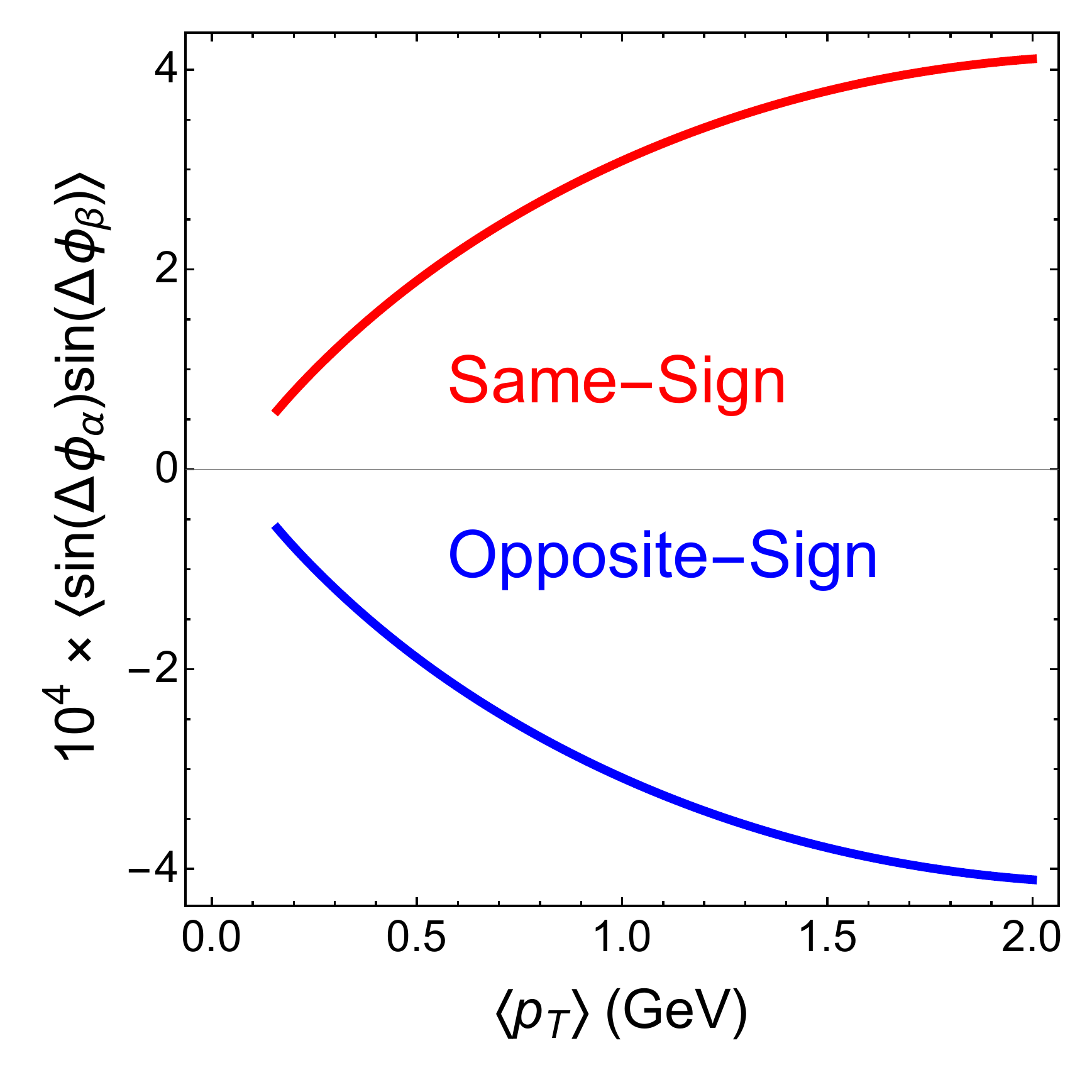} \hspace{0.02in}
\includegraphics[width=0.32\textwidth]{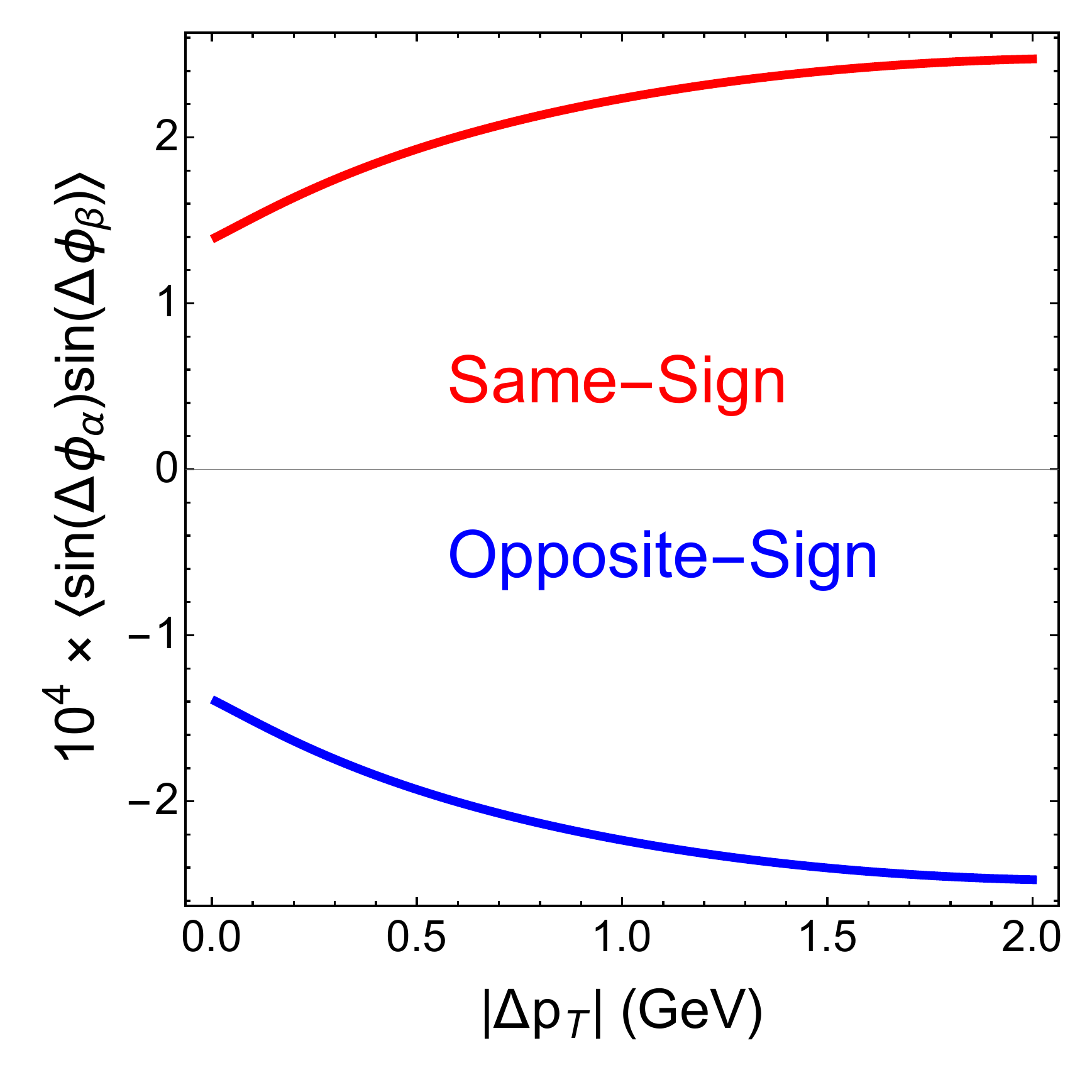}
\caption{(color online) (left) Transverse momentum $p_T$ dependence of charge separation $a_1$;
(middle) Out-of-plane two-particle correlations versus $\left< p_T \right> \equiv \frac{1}{2} (|p_{T,\alpha}|+|p_{T,\beta}|)$;
(right) Out-of-plane two-particle correlations versus $|\Delta p_T| \equiv \Big||p_{T,\alpha}|-|p_{T,\beta}| \Big|$. }  \label{fig_pt}
\end{center}
\end{figure*}

The AVFD simulations can further provide $p_T-$differential information of the CME signal. As shown in Fig.~\ref{fig_pt} left panel (for $50-60\%$ most-central 200GeV Au-Au collisions), the charge separation $a_1$ increases with higher and higher transverse momentum, basically following a hydrodynamic pattern via Cooper-Frye thermal production.  
One can also examine the associated   {out-of-plane} two-particle correlations
\begin{equation}
\left< \sin(\Delta\phi_\alpha) \sin(\Delta\phi_\beta) \right> \equiv \left< \sin(\phi_\alpha-\psi_{EP}) \sin(\phi_\beta-\psi_{EP}) \right>_{\alpha,\beta},
\end{equation}
versus the pair-``averaged'' transverse momentum $\left<p_T\right>$ and the pair-``relative'' transverse momentum $|\Delta p_T|$ defined as 
\begin{eqnarray}
\left< p_T \right> &\equiv& \frac{1}{2} (|p_{T,\alpha}|+|p_{T,\beta}|) ,\\
\left| \Delta p_T \right| &\equiv& \Big||p_{T,\alpha}|-|p_{T,\beta}| \Big|.
\end{eqnarray} 
The results for such dependence are shown in Fig.~\ref{fig_pt} middle and right panels (for $50-60\%$ most-central 200GeV Au-Au collisions). 
While the $\left< p_T \right>$-dependence shows a similar trend as the individual $p_T$, the $| \Delta p_T |$-dependence shows a somewhat flatter trend due the fact that contributions to fixed $| \Delta p_T |$ come from the whole momentum regime. Both results show qualitative agreement  with the experimental measurements \cite{STAR_LPV1,STAR_LPV2,STAR_LPV3}.
 
\section{Predictions for Charge Separation in Isobaric Collisions}\label{sec.isobar}

As previously emphasized, the main challenge for the search of CME in heavy ion collisions is to separate CME-induced signal from background correlations. The difficulty lies in that many sources contribute as backgrounds and currently these contributions are poorly constrained theoretically. Therefore, it is highly desirable to develop more experimentally oriented approach such as new analysis methods or new observables.  Besides the two-component decomposition into H-correlation as discussed above, a number of different proposals were also  put forward, see e.g.~\cite{Voloshin:2004vk,Skokov:2016yrj,Deng:2016knn,Wen:2016zic,Magdy:2017yje,Zhao:2017nfq,Xu:2017qfs}, each with certain advantages. A most promising approach, is to conduct a dedicated isobaric collision experiment, which has now been planned for the 2018 run at RHIC~\cite{Skokov:2016yrj}.  In such ``contrast'' colliding systems (specifically for $^{96}_{40}$Zr$-^{96}_{40}$Zr versus $^{96}_{44}$Ru$-^{96}_{44}$Ru), they have the same baryonic number $A$ but different electric charge $Z$. The expectation is that their bulk evolutions (and thus background correlations) would be basically identical while their magnetic field strength would be different which in turn implies a corresponding difference in the CME signal. For Ru and Zr isobars, there is a $\sim 10\%$ difference in the total charge as well as magnetic field strength: therefore a shift of $\sim20\%$ should be expected for CME-driven correlations on top of identical backgrounds between the two. This will be a crucial test for the search of CME, and quantitative predictions are important. 

\begin{figure*}[!hbt]
\begin{center} 
\includegraphics[width=0.35\textwidth]{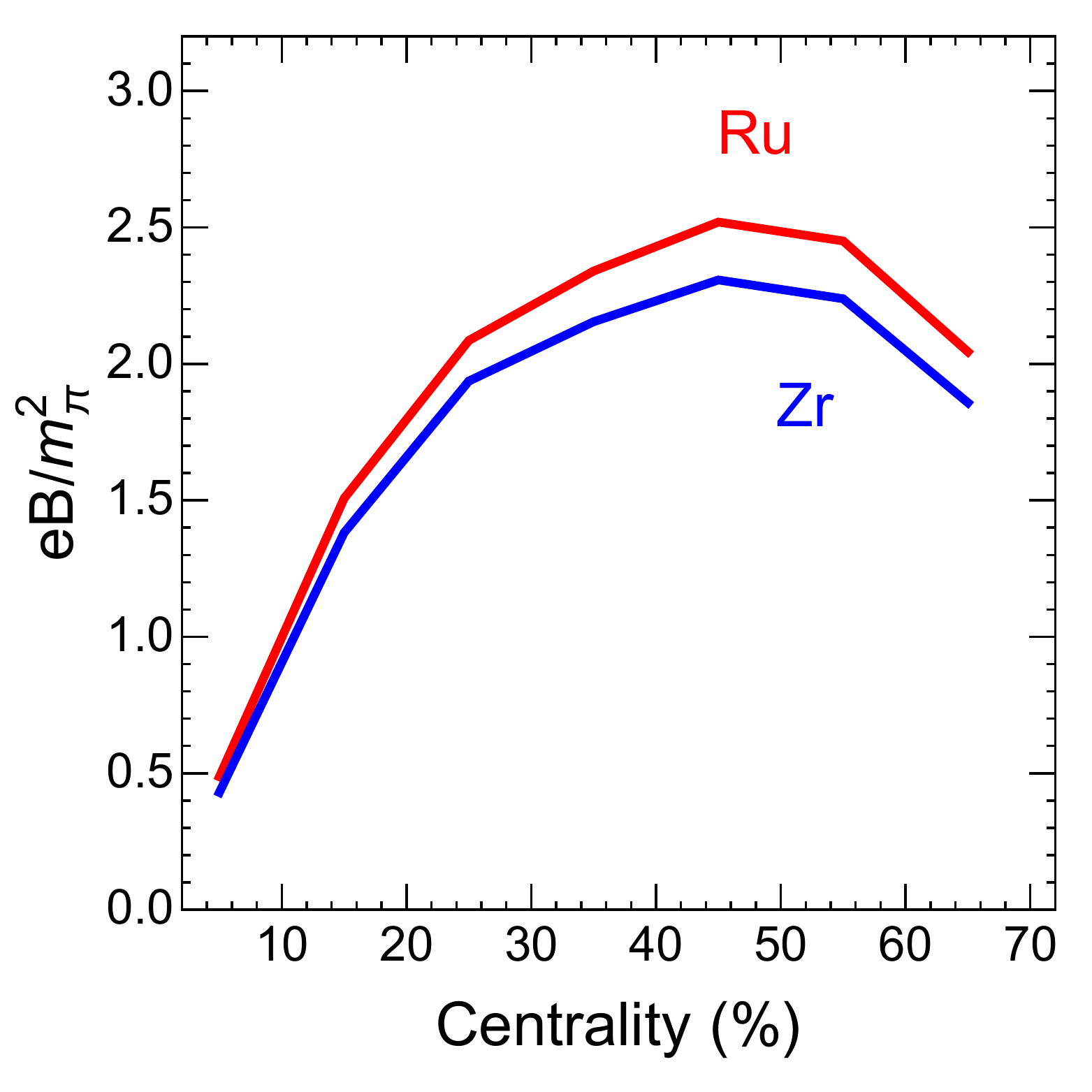}  \hspace{0.04in}
\includegraphics[width=0.35\textwidth]{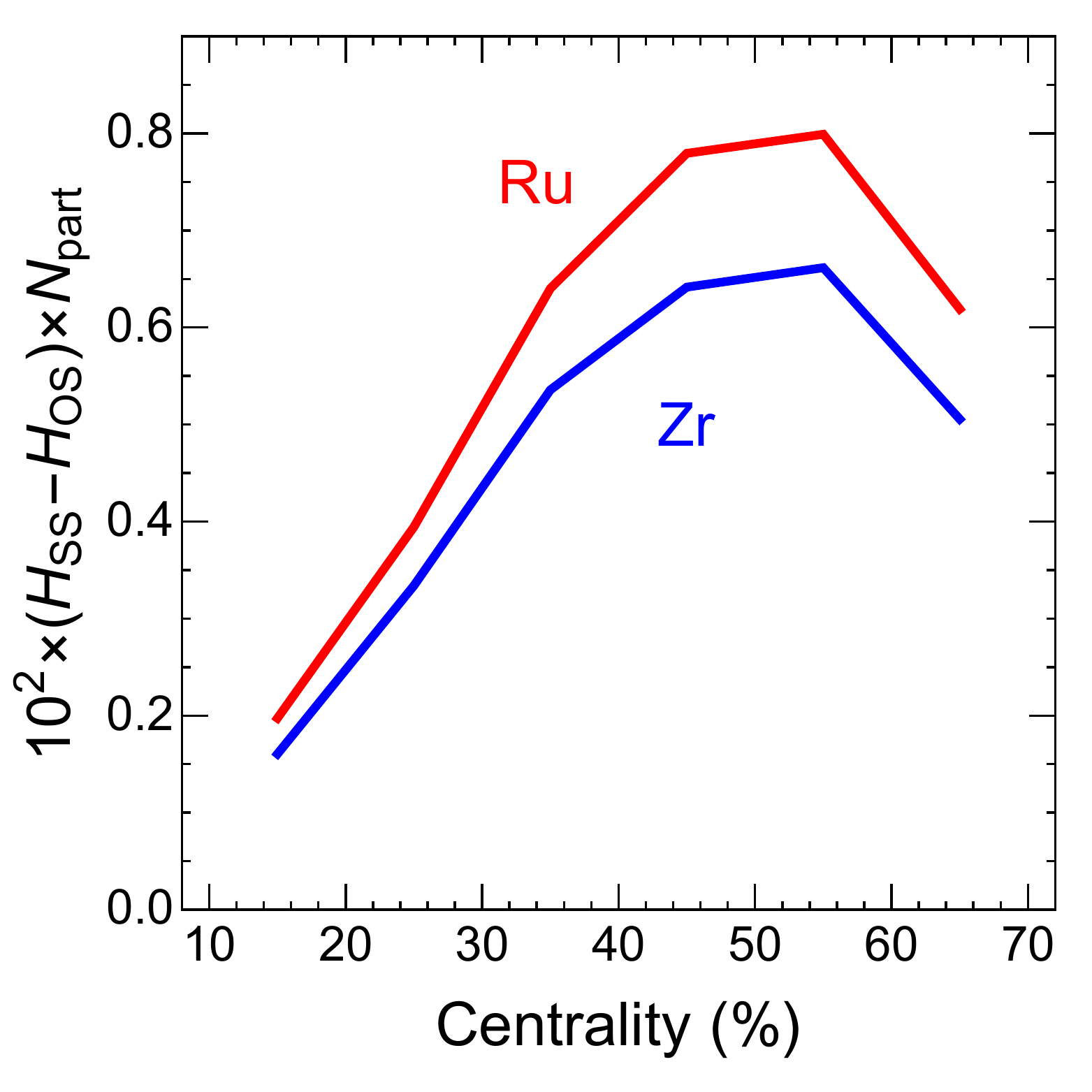} \\
\includegraphics[width=0.35\textwidth]{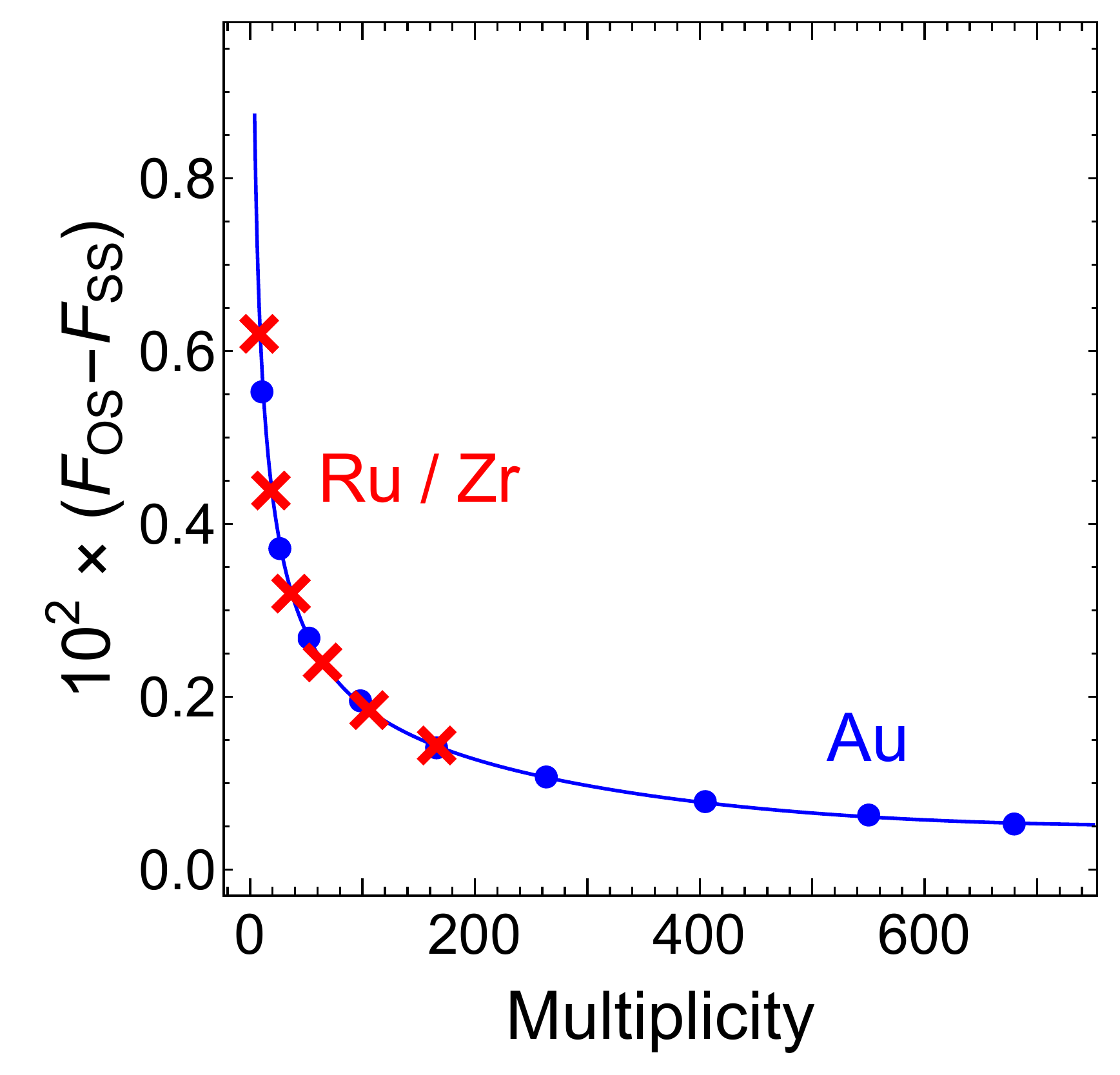} \hspace{0.04in}
\includegraphics[width=0.35\textwidth]{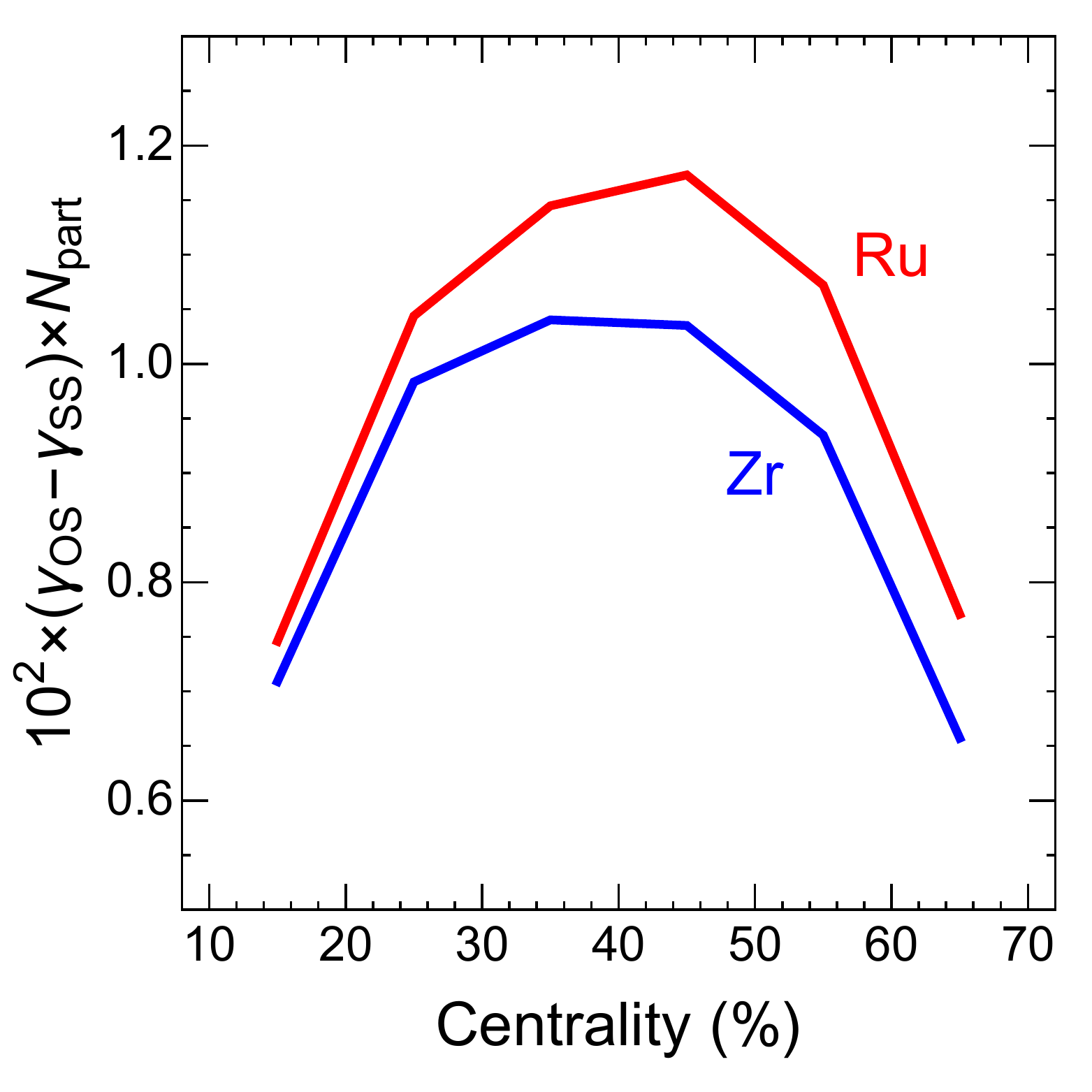}
\caption{(color online) (upper left) Projected initial magnetic field with respect to participant plane given by Monte Carlo Glauber simulation. (upper right) AVFD predictions for CME-induced $H$-correlations in isobar collisions; (lower left) Fitting of background-induced $F$-correlations from Au-Au measurement (blue circle), and the corresponding expectations for Zr-Zr and Ru-Ru systems (red cross); (lower right) Predicted $\gamma$-correlations in Zr-Zr and Ru-Ru collisions by folding together $F$- and $H$-correlations.} \label{fig_isobar}
\end{center}
\end{figure*}
With the AVFD framework developed here for RHIC collisions at 200GeV, we now quantitatively compute the expected CME signal for the isobar colliding systems. Note that the various inputs for AVFD have been fixed via Au-Au collisions and there is no further tuning of parameters, i.e., we take the initial magnetic field as that projected with respect to the participant plane, $B(\tau=0)\equiv\langle B^2 \cos(2\psi_B-2\psi_{2}-\pi) \rangle^{1/2}$, with life time $\tau_B=0.6$fm/c, and $Q_s^2=1.25\mathrm{GeV}^2$ as the ``central curve''. In Fig.~\ref{fig_isobar} (upper right panel), we show the AVFD predictions for pure CME signal $H_{\mathrm{SS}}-H_{\mathrm{OS}}=2(a_1^{ch})^2$ in these two systems. As expected, one can clearly see the $\sim20\%$ difference between them. It is worth emphasizing that the absolute value of H-correlator is sensitive to the chirality imbalance $n_5/s$, determined by saturation scale $Q_s$, the relative difference of H-correlator between Ru-Ru and Zr-Zr systems is independent.

On the other hand, as such ``pure'' signal $H$ is not measured directly in experiments, one would be interested in also computing the directly measured correlation $\gamma_{\mathrm{OS}}-\gamma_{\mathrm{SS}}$, and examining whether there would be sizable difference in $\gamma_{\mathrm{OS}}-\gamma_{\mathrm{SS}}$ between the isobaric systems. To do so, however, requires knowledge about the non-CME background contributions. Our strategy here is to extrapolate the likely background level to the isobar collisions, based on the $F+H$ two-component decomposition and the assumption that $F$ is mainly a function of multiplicity only~\cite{Bloczynski:2013mca}.  From the results of RHIC $\sqrt{s_{NN}}=200$GeV Au-Au collisions \cite{STAR_LPV_BES}, one obtains the background $F_{\mathrm{OS}}-F_{\mathrm{SS}}$ versus the measured charged particle multiplicities, shown as blue circles in Fig.~\ref{fig_isobar}(lower left panel). We have also performed a fitting of the $F_{\mathrm{OS}}-F_{\mathrm{SS}}$ versus multiplicity (for Au-Au points)  with an algebraic-fractional-formula $F(x)=\frac{1\,+\,b_1\,x\,+\,b_2\,x^2 }{c_1\,x\,+\,c_2\,x^2}$, shown as the blue curve. (The best fit is given by $b_1=0.119$, $b_2=5.81\times10^{-5}$, $c_1=33.7$, and $c_2=0.395$.) 
Given the fitting curve, one could then read a plausible estimate of the expected $F$-correlations in correspondence to the multiplicity in Ru-Ru and Zr-Zr systems, shown as the red crosses in the middle panel of Fig.~\ref{fig_isobar} for centrality bins from $10-20\%$ (with higher multiplicity) to $60-70\%$ (with lower multiplicity).  

\begin{table}[!hbt]\centering
\begin{tabular}{ccccccc}
\hline
centrality bin & 10-20\% & 20-30\% & 30-40\% & 40-50\% & 50-60\% & 60-70\% \\
\hline
$N_{part}$ & 100.46 & 67.17 & 43.04 & 25.79 & 14.06 & 6.81  \\
\hline
Multiplicity & 167.0 & 106.9 & 65.43 & 37.43 & 19.52 & 9.096 \\
\hline
$eB_0[\mathrm{Ru}](m_\pi^2)$ & 1.507 & 2.086 & 2.340 & 2.520 & 2.450 & 2.044 \\
\hline
$eB_0[\mathrm{Zr}](m_\pi^2)$ & 1.381 & 1.937 & 2.155 & 2.307 & 2.239 & 1.858 \\
\hline
$n_5/s$ & 0.097 & 0.114 & 0.144 & 0.188 & 0.266 & 0.394  \\
\hline
$10^4\times(H_{\mathrm{OS}}-H_{\mathrm{SS}})[\mathrm{Ru}]$ & 0.196 & 0.587 & 1.487 & 3.023 & 5.682 & 9.091 \\
\hline
$10^4\times(H_{\mathrm{OS}}-H_{\mathrm{SS}})[\mathrm{Zr}]$ & 0.160 & 0.497 & 1.245 & 2.488 & 4.705& 7.425 \\
\hline
$10^3\times(F_{\mathrm{OS}}-F_{\mathrm{SS}})$ & 1.42 & 1.83 & 2.38 & 3.18 & 4.37 & 6.19 \\
\hline
\end{tabular}
\caption{Centrality dependence of participant number, multiplicity, B field, axial charge, computed CME signal H, and extrapolated non-CME background F in isobar system. The $n_5/s$ and corresponding $H_{\mathrm{OS-SS}}$ correlators shown here are obtained with a  saturation scale $Q_s^2=1.25{\rm GeV}^2$.\label{tab2}}
\end{table}

With both the AVFD-predicted CME signal $H$-correlations and the extrapolated bulk background $F$-correlations (see Tab.\ref{tab2}), one can then make a prediction for the $\gamma$-correlations via Eq.~\ref{eq_cme_correlation}. The results are shown in Fig.~\ref{fig_isobar}(lower right panel), there is a visible $\sim 15\%$ relative difference between the two systems for the $\gamma$-correlations in the relatively peripheral collisions. Provided the present projections for  uncertainty and limitations in measurements~\cite{Skokov:2016yrj,Deng:2016knn}, a $15\%$-level difference should be readily detectable at the scheduled isobaric collision experiment.

\section{Possible Relaxation Effect for the Anomalous Transport Currents}\label{sec.relaxative}

As briefly mentioned in Sec.~\ref{subsec.avfd}, the CME current is treated without any relaxation effect in the ``standard'' AVFD framework, which assumes that the CME current establishes instantaneously in response to the magnetic field and chirality imbalance. In other words, the time needed for the hot medium to respond to the changing external field or axial charge, is assumed to be negligible as compared with relevant QGP evolution time scales. This is plausible as the CME current is a non-dissipative quantum transport current that should occur on microscopic quantum evolution time scale which usually is negligibly short compared with any macroscopic scale.  Theoretically this is a subtle issue that has not been fully settled. A recent discussion in \cite{Kharzeev:2016sut} suggests that a fraction of the total CME current may be subject to thermal relaxation effect.  In such situation, it would be crucial to calibrate the influence of such theoretical uncertainty. 

In order to do this, the AVFD framework needs to be slightly adapted to treat a proper fraction of the CME current in a similar fashion as the Navier-Stokes current, {\it i.e.} the relaxation effect should be included via the second order viscous terms. In this section, we investigate this uncertainty by comparing two extreme scenarios: the ``instant-CME" as in Eqs.(\ref{eq_avfd}-\ref{eq_avfd_ns_2}), versus the $100\%$-relaxation scenario below:
\begin{eqnarray} 
\hat{D}_\mu J_{\chi, f}^\mu &=& \chi \frac{N_c Q_f^2}{4\pi^2} E_\mu B^\mu \\
J_{\chi, f}^\mu &=& n_{\chi, f}\, u^\mu + \nu_{\chi, f}^\mu  \\ 
\Delta^{\mu}_{\,\, \nu} \hat{d} \left(\nu_{\chi, f}^\nu \right) &=& - \frac{1}{\tau_{r}} \left[  \left( \nu_{\chi, f}^\mu \right) -  \left(\nu_{\chi, f}^\mu \right)_{NS} \right ] \\
\left(\nu_{\chi, f}^\mu \right)_{NS} &=&  \frac{\sigma}{2} T \Delta^{\mu\nu}   \partial_\nu \left(\frac{\mu_{\chi, f}}{T}\right) +  \frac{\sigma}{2} Q_f E^\mu + \chi \frac{N_c Q_f}{4\pi^2} \mu_{\chi, f} B^\mu
\end{eqnarray} 

\begin{figure}[!hbt]
\begin{center}
\includegraphics[width=0.5\textwidth]{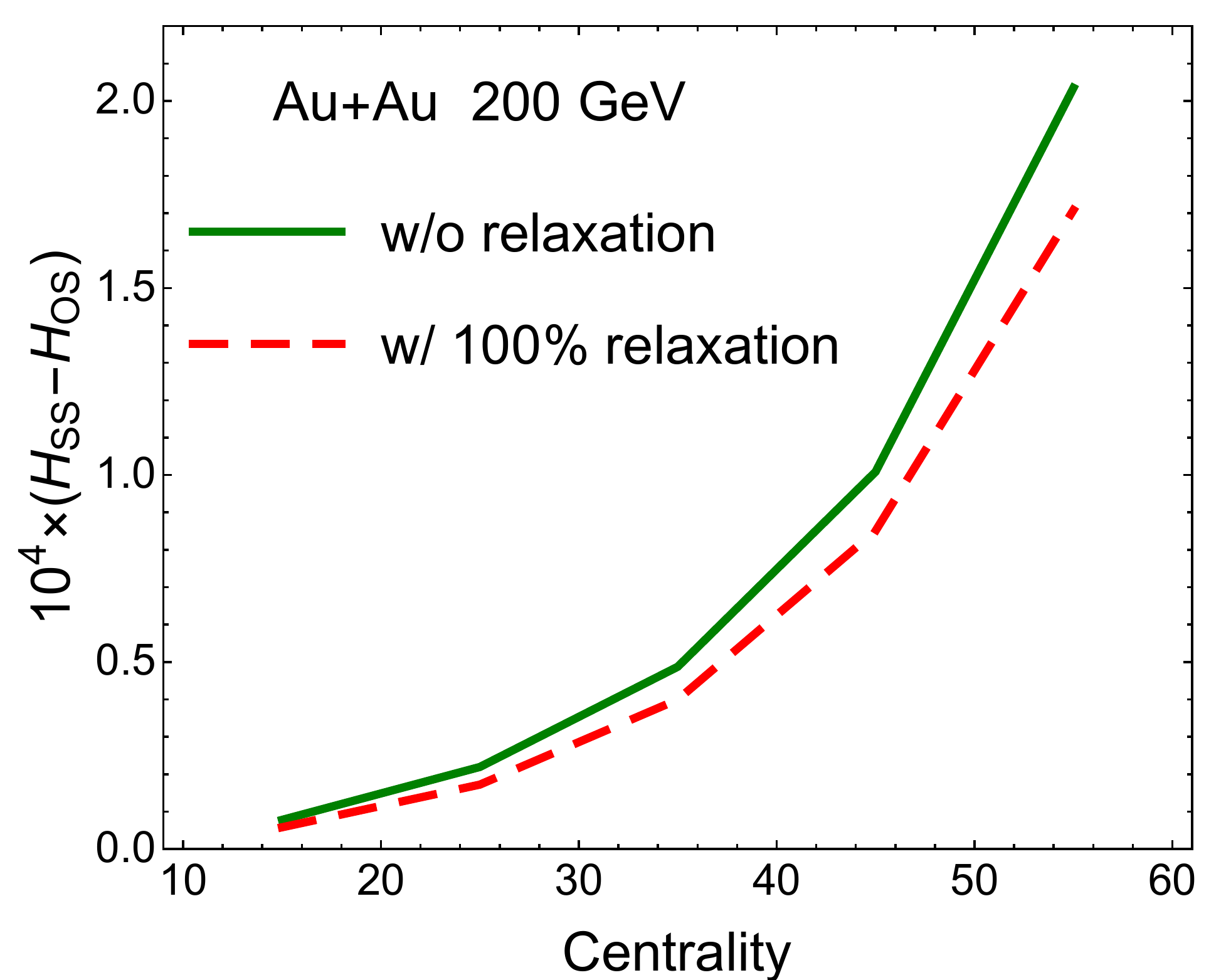}  
\caption{(color online) Comparison of two AVFD calculations with different scenarios for the relaxation effect on the CME current: the ``instant-CME" scenario (solid green line) without relaxation versus the $100\%$-relaxation scenario (dashed red line).}
\label{fig_relaxativeCME}
\end{center}
\end{figure} 

In Fig.~\ref{fig_relaxativeCME} we show such a comparison to demonstrate the influence of the potential relaxation effect on the CME current. The results suggest a very mild sensitivity, with the instant-CME scenario giving a stronger charge separation signal. The relaxation effect has two opposite impacts on the CME current: it delays the buildup of the CME current at the beginning, while on the other hand also delays the decay of the CME current (once it is there) with decreasing magnetic field. Due to the competition of the two, even this extreme scenario with $100\%$-relaxation only reduces the signal mildly. The realistic case should be somewhere in between these two curves.

It is worth emphasizing that in order to see the influence of potential relaxation effect, we start both scenarios, with or without relaxation effect, with exactly the same initial condition of axial and vector charge density, but no initial axial or vector currents, i.e. assuming no pre-thermal CME. One can expect that if with the pre-thermal CME current, the relaxation effect would slowdown the decay of the CME current, and help to develop more charge separation.

\section{Finite Quark Mass and Strangeness Transport}\label{sec.strangeness}

Another question of both theoretical interest and experimental relevance, is whether the strange quarks and antiquarks undergo the anomalous transport and contribute to certain observables. For the $u$ and $d$ quarks in the QGP, their masses are negligibly small compared with temperature scale, therefore they can be well approximated as chiral fermions. This is not the case for the strange quark or antiquark, whose mass is not small compared with temperature, $\frac{M_s}{T} \sim \frac{1}{3}$. A finite mass will lead to random flipping of chirality and cause dissipation of nonzero axial charge. As a result, the anomalous transport will be suppressed. It has been recently argued that such finite mass effect scales as $\left(\frac{M_s}{T} \right)^2$~\cite{Guo:2016nnq}, which in the case of strange quark would not be a severe suppression. This would thus leave certain room for  potential contributions to anomalous transport processes from strangeness sector. 

To get some quantitative insight on the impact of this issue, let us consider two extreme cases: the case with strange quark experiencing  the anomalous transport, equally as the light quarks (aka. ``3-flavor'' case); or the case with no anomalous transport at all for $s$ quarks (aka. ``2-flavor'' case). 
\begin{figure}[!hbt]
\begin{center}
\includegraphics[width=0.4\textwidth]{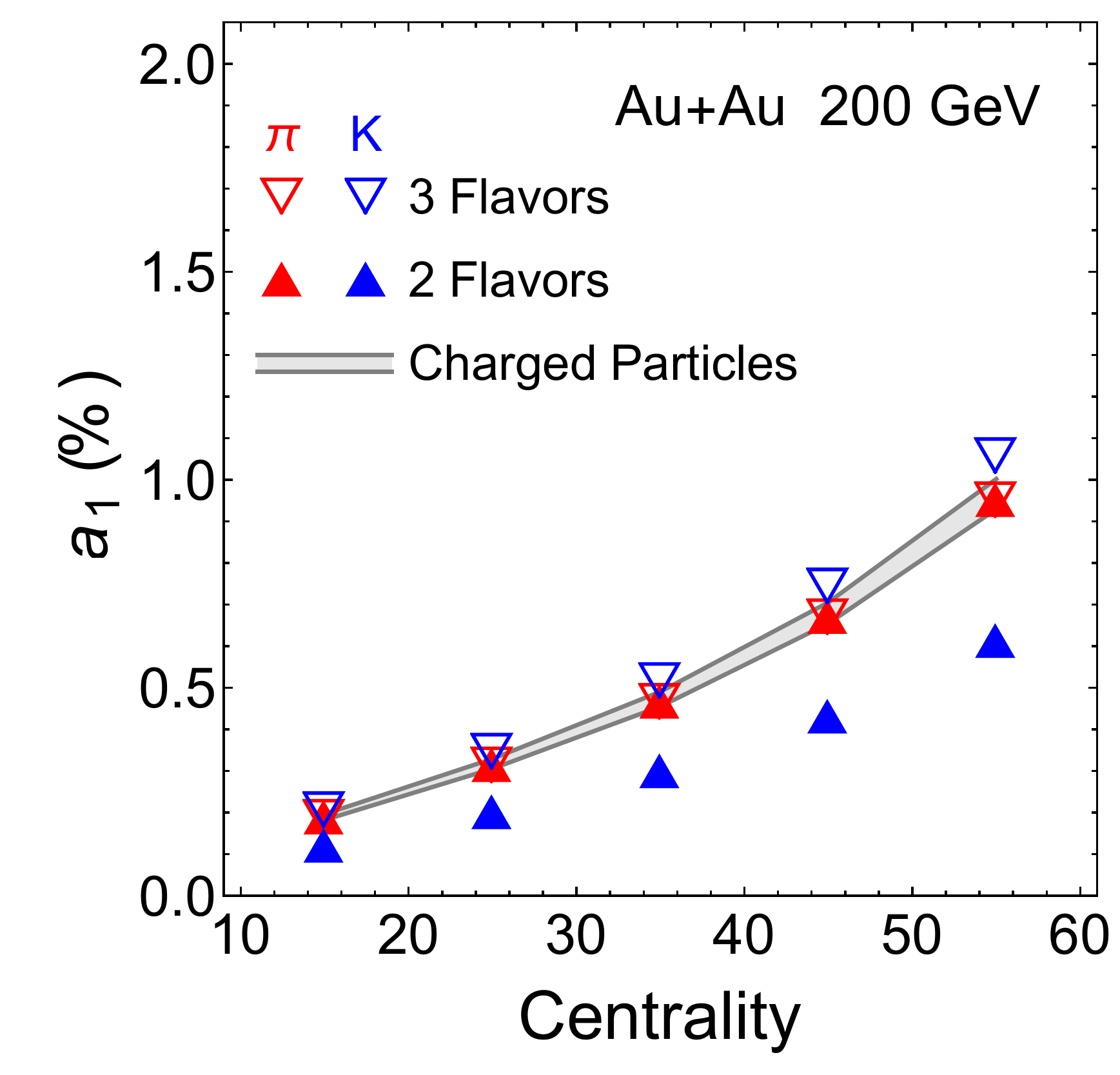}  \quad
\includegraphics[width=0.4\textwidth]{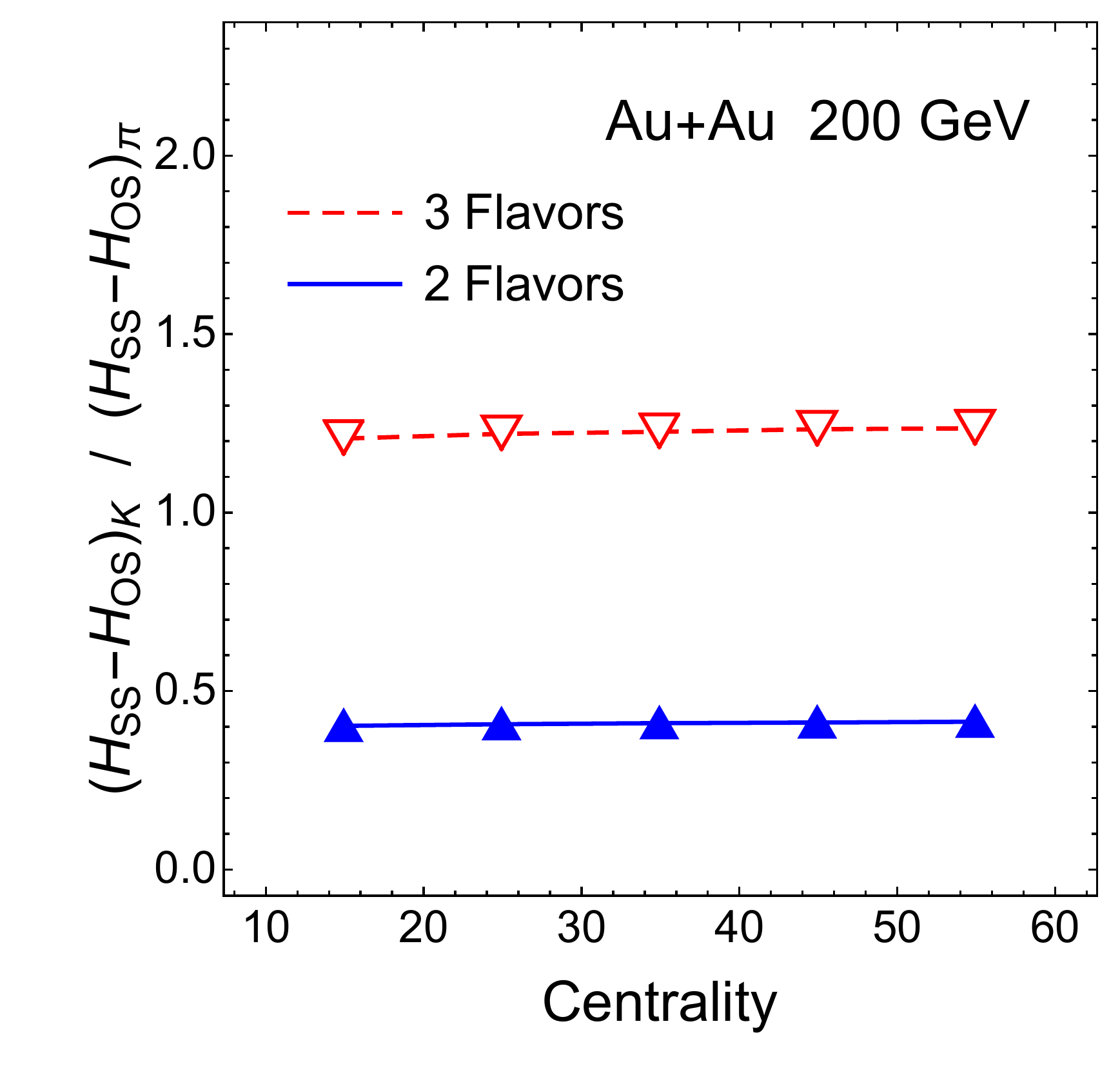}  
\caption{(color online) CME signal of identified particles for the ``2-flavor case'' (upward solid triangles) versus the ``3-flavor case''  (downward open triangles). 
(left) Charge separation $a_1$ of charged pions   (red symbols) and charged kaons (red symbols) for different centrality, with the gray band for the charge separation $a_1$ of all charged particles. 
(right) The corresponding ratio of $K^{\pm}$ H-correlations  
 to  $\pi^{\pm}$ H-correlations versus centrality. } 
\label{fig_strangeness}
\end{center}
\end{figure} 

From Fig.~\ref{fig_strangeness} (left panel) one can find that the charge separation signal of either pions only or all charged particles  is insensitive to strangeness contributions, whereas the  charge separation signal of kaons is extremely sensitive to any strangeness contributions. On the right panel we show the ratio of $K^{\pm}$ H-correlations  
 to  $\pi^{\pm}$ H-correlations: the reality could well lie in between these two extreme scenarios, and a precise experimental measurement of this ratio will provide highly valuable insight into the anomalous transport of the strangeness sector. 

\section{Discussions on the Evolution of Pre-Hydro Charge Separation}\label{sec.prehydro}

Finally, we discuss the potential contribution from the pre-hydro charge separation. As discussed before, the magnetic field is very strong at the early time, even before the start time of hydrodynamic evolution. The CME is a general transport phenomenon that would occur both in the equilibrium and the out-of-equilibrium setting. It is therefore conceivable that there could already be CME-induced charge separation during the pre-hydro stage. Indeed there have been various studies of pre-equilibrium generation of charge separation, see e.g. \cite{Mace:2016shq,Mace:2016svc,Mueller:2016ven,Fukushima:2015tza,Sun:2016nig,Huang:2017tsq,Ma:2011uma}. This implies that by the start of hydrodynamics, the density and current could already become nontrivial.  

Such pre-hydro charge separation can be naturally integrated with AVFD framework as initial conditions for the corresponding fermion densities and currents at the hydro initial time $\tau_0$. In the present section, we  investigate how the nontrivial initial conditions via  charge density  dipole and charge current along $\vec{\bf B}$ field from the pre-hydro evolution will  propagate through the hydrodynamic stage toward final hadron observables. To do this, 
 one can recast RH/LH currents into vector/axial currents and rewrite Eqs.(\ref{eq_avfd}-\ref{eq_avfd_ns_2}) as  
\begin{eqnarray}
\hat{D}_\mu J_{f}^\mu &=& 0 \\
J_{f}^\mu &=& n_{f}\, u^\mu + \nu_{f}^\mu \\
\Delta^{\mu}_{\,\, \nu} \hat{d} \left(\nu_{f}^\nu \right) &=& - \frac{1}{\tau_{r}} \left[  \left( \nu_{f}^\mu \right) -  \left(\nu_{f}^\mu \right)_{NS} \right ] \\
\left(\nu_{f}^\mu \right)_{NS} &=&  \sigma T \Delta^{\mu\nu} \partial_\nu \left(\frac{\mu_{f}}{T}\right)\quad
\end{eqnarray}
and similar equations can be obtained for $J_5, n_5$.

\begin{figure}[!hbt]
\begin{center} 
\includegraphics[width=0.32\textwidth]{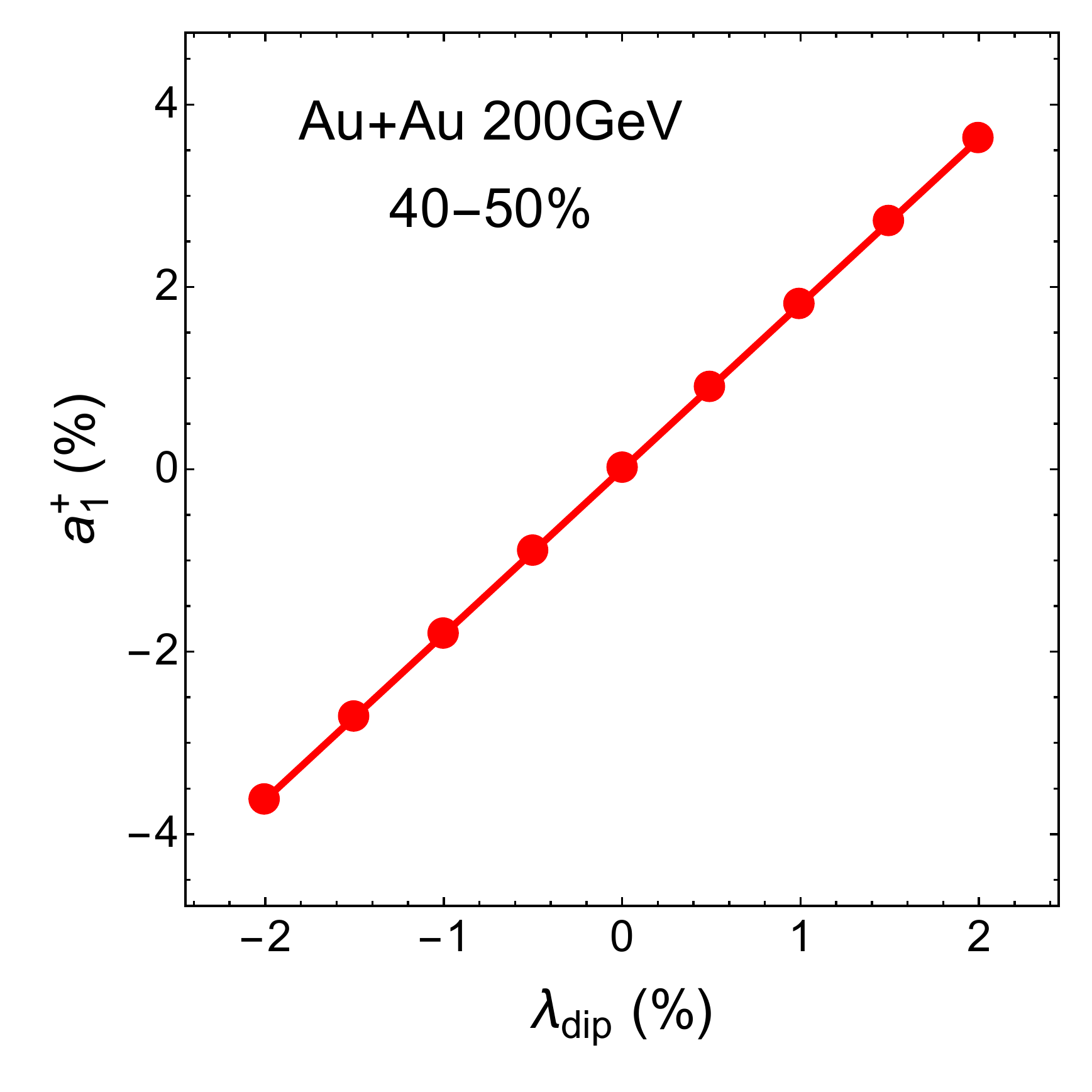}  
\includegraphics[width=0.32\textwidth]{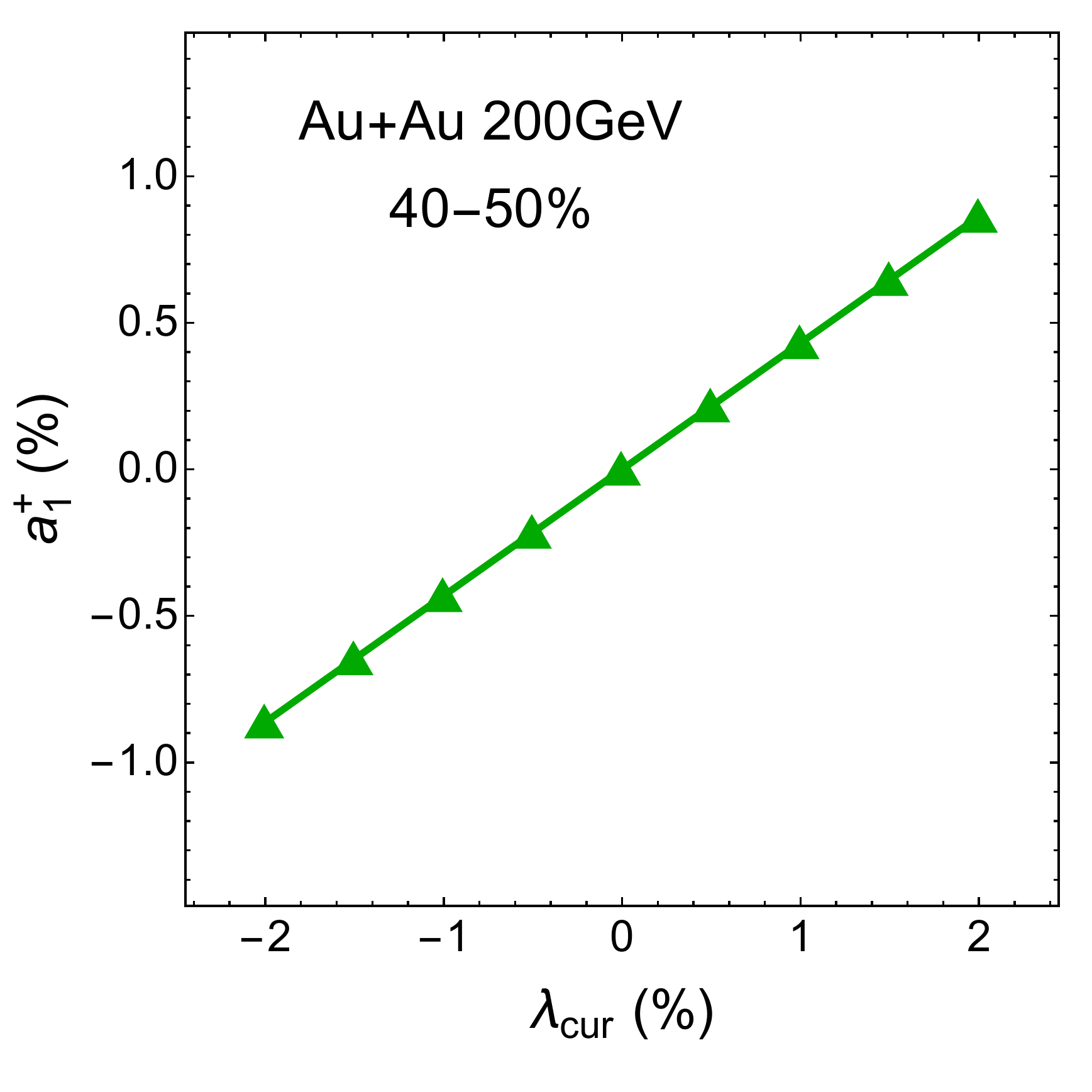}  
\includegraphics[width=0.32\textwidth]{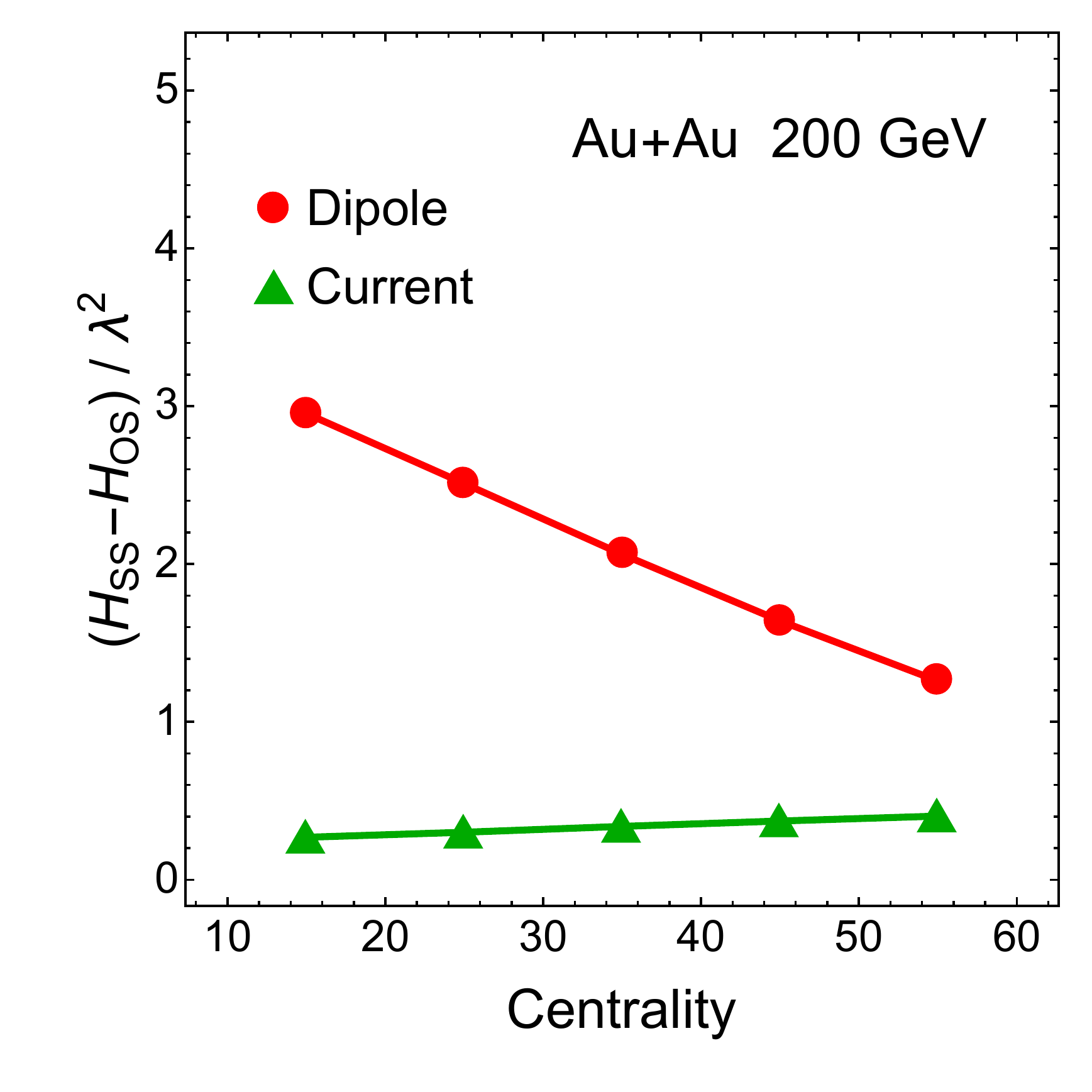}  
\caption{(color online) Final charge separation signal computed via AVFD with nontrivial initial conditions from  pre-hydro CME-induced charge dipole or current:  
(left) Charge separation $a_1$ caused by nonzero initial dipole;  
(middle) Charge separation $a_1$ caused by nonzero initial current; 
(right) Centrality dependence of $H^{OS-SS}$ correlations due to nonzero initial dipoles (red) or currents (green), normalized by the respective initial condition parameters. }
\label{fig_prehydro}
\end{center}
\end{figure}

To quantify the pre-hydro currents and charge dipole, we rescale them by the initial entropy density $s_0$ at $\tau_0$, with dimensionless factor $\lambda$.
When solving the propagation of initial currents, we start with the initial condition that
\begin{eqnarray}
\nu_{f}|_{\tau=\tau_0} = J_{\mathrm{ini},f}= \lambda_{\mathrm{cur},f}\; s_0 \;\hat{y},
\end{eqnarray}
while for initial charge dipole,
\begin{eqnarray}
n_{\mathrm{ini},f} = (\lambda_{0,f}+\lambda_{\mathrm{dip},f} \sin\phi)\;s_0.
\end{eqnarray}
Also, as these pre-hydro currents/dipoles are due to the Chiral Magnetic Effect, one could expect that they should be proportional to the corresponding quarks' electric charge:
\begin{eqnarray}
\lambda_{\mathrm{dip}} \equiv \lambda_{\mathrm{dip},u} = -2\lambda_{\mathrm{dip},d},\\
\lambda_{\mathrm{cur}} \equiv \lambda_{\mathrm{cur},u} = -2\lambda_{\mathrm{cur},d}.
\end{eqnarray}
Starting from such initial conditions with given $\lambda_{\mathrm{dip}}$ or $\lambda_{\mathrm{cur}}$, we solve the AVFD equations and compute the final state hadron charge separation signal $a_1$. Fig.~\ref{fig_prehydro} shows the results for $a_1$ due to {\sl only} pre-hydro charge dipole (left panel) or current (middle panel) for given initial condition parameters for a particular centrality. One finds that the final signal responds linearly to the initial charge separation from pre-hydro CME. In the right panel we show the final correlations $H^{OS-SS}$ scaled by $\lambda_{\mathrm{dip}}$ or $\lambda_{\mathrm{cur}}$ for the two types of initial conditions for a variety of centrality.  
The results suggest that the pre-hydro charge separation could reach a level at several percent of the initial entropy density, its effect on final signal would contribute a substantial fraction of the experimental data (around the magnitude of $\sim10^{-4}$). In the future, when pre-equilibrium models could quantitatively compute the CME-induced early charge separation, one could then use those results as input for AVFD to predict final observables. 

\section{Conclusion}\label{sec.conclusion}

In this paper, we report a detailed study of the recently developed Anomalous-Viscous Fluid Dynamics (AVFD) framework for quantifying the charge separation signal from the Chiral Magnetic Effect in heavy ion collisions. The AVFD incorporates both normal viscous transport effects and the anomaly induced transport effect. It solves the evolution equations of the fermion currents in QGP, on top of the neutral bulk background described by data-valid VISH2+1 hydrodynamic simulation. With such tool, we quantitatively investigate the sensitivity of CME signal to a series of key parameters, including the time dependence of the magnetic field, the initial axial charge, the viscous transport coefficients as well as the resonance decay contributions. With realistic initial conditions and magnetic field lifetime, the predicted CME signal is quantitatively consistent with measured charge separation data in 200GeV Au-Au collisions. We further predict the CME observables for the upcoming isobaric (Ru-Ru v.s. Zr-Zr ) collision experiment  that could provide the critical test for the presence of the CME.

The AVFD framework has further allowed us to investigate for the first time the influence on the CME signal by several  theoretical uncertainties. We find that the possible thermal relaxation effect on the CME transport current  has a mild impact on the final signal. The potential contribution to anomalous transport from the strangeness sector could have a substantial and observable consequence on the charged kaon correlation signals. In addition with the AVFD we quantify the final state charge separation arising entirely from nontrivial charge or current initial conditions due to the pre-hydro CME contributions.  

A number of future developments of the AVFD framework is underway, including: the event-by-event AVFD simulations; the direct implementation of background effects into the formalism; as well as the quantification of another important anomalous transport phenomenon, namely the Chiral Magnetic Wave which would lead to an observable elliptic flow difference of positively and negatively charged particles~\cite{Burnier:2011bf}. These investigations will be reported in a forthcoming publication.

\vspace{0.2in}

 {\bf Acknowledgments.} 
The authors thank D. Hou, D. Kharzeev, M. Stephanov, H.-U. Yee and P. Zhuang 
for helpful discussions. The authors are particularly grateful to Y. Yin for his collaboration at the early stage of this project and to U. Heinz and C. Shen for developing the VISH2+1 hydro simulations and for their assistance in using this tool. This material is based upon work supported by the U.S. Department of Energy, Office of Science, Office of Nuclear Physics, within the framework of the Beam Energy Scan Theory (BEST) Topical Collaboration. JL and SS are also partly supported by the National Science Foundation under Grant No. PHY-1352368. YJ is supported by the Beijing University under Startup Funding. EL acknowledges support from the NSF REU program grant PHY-1460882. The computation of this research was performed on IU's Big Red II \& Karst clusters,
that are supported in part by Lilly Endowment, Inc., through its support for the Indiana University Pervasive Technology Institute, and in part by the Indiana METACyt Initiative. The Indiana METACyt Initiative at IU was also supported in part by Lilly Endowment, Inc.

\end{document}